\documentclass[a4paper, 11pt]{article}

\usepackage{graphicx}%
\usepackage{multirow}%
\usepackage{amsmath,amssymb,amsfonts}%
\usepackage{amsthm}%
\usepackage{mathrsfs}%
\usepackage[title]{appendix}%
\usepackage{xcolor}%
\usepackage{textcomp}%
\usepackage{manyfoot}%
\usepackage{booktabs}%
\usepackage{algorithm}%
\usepackage{algorithmicx}%
\usepackage{algpseudocode}%
\usepackage{listings}%
\usepackage{natbib}
\usepackage{rotating}

\usepackage{caption}
\usepackage[a4paper,top=3cm,bottom=4cm,left=3cm,right=3cm,marginparwidth=1.75cm]{geometry}

\usepackage[colorlinks=true, allcolors=blue]{hyperref}

\title{Multivariate Bayesian models with flexible shared interactions for analyzing spatio-temporal patterns of rare cancers}
\author{Garazi Retegui$^{1,2}$, Jaione Etxeberria$^{1,2}$, Mar{\'i}a Dolores Ugarte$^{1,2*}$\\
	\\
	\small $^1$ Department of Statistics, Computer Science and Mathematics, Public University of \\
	\small Navarre (UPNA), Arrosadia Campus, Pamplona, 31006, Navarra, Spain.\\
	\small $^2$ Institute for Advanced Materials and Mathematics (INAMAT2), Public University \\ 
	\small of Navarre (UPNA), Arrosadia Campus, Pamplona, 31006, Navarra, Spain.\\
	\small $^{*}$ Corresponding author.\\
	\\
	\small lola@unavarra.es
	\date{}
}

\begin{document}
	\maketitle
	
	\begin{abstract}
		\noindent Rare cancers affect millions of people worldwide each year. However, estimating incidence or mortality rates associated with rare cancers presents important difficulties  and poses new statistical methodological challenges. In this paper, we expand the collection of multivariate spatio-temporal models by introducing adaptable shared spatio-temporal components to enable a comprehensive analysis of both incidence and cancer mortality in rare cancer cases. These models allow the modulation of spatio-temporal effects between incidence and mortality, allowing for changes in their relationship over time. The new models have been implemented in INLA using r-generic constructions. We conduct a simulation study to evaluate the performance of the new spatio-temporal models. Our results show that multivariate spatio-temporal models incorporating a flexible shared spatio-temporal term outperform conventional multivariate spatio-temporal models that include specific spatio-temporal effects for each health outcome. We use these models to analyze incidence and mortality data for pancreatic cancer and leukaemia among males across 142 administrative health care districts of Great Britain over a span of nine biennial periods (2002-2019).
	\end{abstract}

	\section{Introduction}\label{Section1:Introduction}
	
	Rare cancers attract interest within the scientific, clinical, and public health community as they represent a significant burden worldwide, affecting millions of people each year. According to the National Cancer Institute in the United States, rare cancers account for approximately 27\% of all cancer diagnoses and 25\% deaths in the country. This translates to more than 700,000 new cases of rare cancers each year \citep{botta2020}. In Europe, rare cancers represent around 24\%  of all cancer diagnoses between 2000 and 2007, with an estimated annual count of over 650,000 newly diagnosed cases \citep{gatta2017}.
	The task of measuring the burden of rare cancers is complex due to the limited availability of comprehensive data and the varying definitions of rare cancers across different countries and organizations. As a result, rare cancers are frequently under-researched. Estimating the incidence and mortality of rare cancers is crucial for several reasons. First, it helps enhance our understanding of the disease's underlying biology and risk factors. Second, it assists public health authorities in allocating resources effectively and identifying areas that require priority action in the case of lethal cancers. Third, it facilitates the design of effective clinical trials. Lastly, it helps advocate for patients and encourages investments in research and treatment \citep{mallone2013, botta2018, salmeron2022}.
	
	Different European initiatives and projects have been carried out in recent years to obtain estimates of incidence, survival, and mortality of rare cancers and even to encourage research, e.g. the EU Joint Action on Rare Cancers (JARC) or the project Surveillance of Rare Cancers in Europe (RARECARE) (\url{https://www.rarecarenet.eu/}).
	Specifically, RARECARE collected data on cancers from 89 population-based cancer registries in 21 European countries, allowing to study the epidemiology of these cancers as a whole in a large and heterogeneous population \citep{gatta2011}. Thus, it fulfills the objective of developing a clinical database on very rare cancers to provide new knowledge on these diseases and to enable updated indicators of rare cancer burden. However, developing new statistical methods to provide updated indicators of the burden of rare cancers is not a specific objective of RARECARE. As a result, the research conducted has used simple techniques. For example, \cite{gatta2011, gatta2017} and \cite{botta2020} estimate incidence rates as the number of new cases occurring in a given period divided by the total person-years in the general population, i.e. they calculate crude incidence rates per \mbox{100 000} inhabitants.  \cite{botta2018} employ a model-based approach using a simple Poisson random-effects model to estimate cancer incidence and thus obtain the yearly expected number of cases for each rare cancer in each European country. However, this study only provides incidence estimates for large areas such as countries, and therefore, it does not provide estimates at a sub-national level using local or regional cancer registries. This limitation poses a challenge in investigating the aetiology of rare cancers in small areas. Additionally, the model used by \cite{botta2018} does not consider the temporal dimension. Examining the geographical pattern of rare cancers over time in small areas within a country provides valuable information for epidemiologists and health researchers to go further in formulating aetiological hypotheses. However, this would need the use of advanced statistical techniques, like the spatio-temporal models used in disease mapping.
	Models incorporating spatial dependence, i.e. borrowing information from neighbouring areas, such as random effects models where the regional effect is modelled using an intrinsic conditional autoregressive models \citep[iCAR,][]{besag1974} or the well-known Besag-York-Molli{\'e} model \citep[BYM,][]{besag1991}, continue to be widely used. These models have been extended to include temporal random effects and spatio-temporal interactions. See for example \cite{goicoa2016} or \cite{carroll2019}.
	
	Analyzing the geographical pattern of a single disease and how it changes over time is highly valuable in identifying potential risk factors that may contribute to the knowledge of the disease. However, this analysis may be limited when the disease is particularly rare. In that case it could be useful to consider additional diseases or health outcomes of the same disease to increase the effective sample size and to be able to analyze the possible (and in many cases unknown) relationships among diseases. The joint modeling of multiple cancer outcomes can be highly beneficial as it allows for the exploration of geographical and temporal patterns while considering potential relationships between them.
	
	There is a considerable amount of theoretical research about multivariate disease mapping models. One approach to joint modelling involves using the multivariate conditional autoregressive models (MCAR). Seminal work by \cite{mardia1988} established a theoretical framework in this domain, extending the seminal work of \cite{besag1974}. Based on the work of \cite{mardia1988}, various proposals can be found in the literature \citep[see for example ][]{gelfand2003, jin2005, jin2007, macnab2011}. 
	A general coregionalization framework for multivariate areal models that covers many of the previous proposals was introduced by \cite{martinez2013}. Nevertheless, similar to most multivariate areal models, this approach may seriously increase computational burden, which makes simultaneous modeling of a moderate to large number of responses unapproachable. \cite{botella2015} presents an interesting alternative to address this issue, the so-called M-based models. These models offer a simpler and computationally efficient technique that achieves a balance between computational tractability and model identification. These models have been employed to explore the spatial correlation among health outcomes, particularly when these correlations are assumed to be initially unknown. On the other hand, a standard and computationally simpler method of extending univariate spatial distributions to the multivariate case is through spatial factor modelling \citep{wang2003}. In such models, the diseases are known to share one or more underlying, unobserved common spatial factors, which are estimated jointly with some loadings weighting their contributions to the geographical pattern of each disease. Shared component models \citep{held2005} can be considered as a special case of the spatial factor model. In this study, we exploit the correlation between incidence and mortality rates among different cancer locations and hence, we consider a set of shared component models. This approach facilitates the identification of shared risk factors among different outcomes and enhances our understanding of the disease's aetiology \citep[see for example ][]{held2006, kazembe2015, retegui2021}. In addition, when the disease being studied has low incidence, shared component models that examine related diseases or health outcomes help to improve estimates by borrowing information from nearby areas or time points \citep{etxeberria2018,retegui2021}.  Therefore, to enhance the accuracy of rare cancer rate estimates in small areas over time, we use multivariate models that simultaneously analyze cancer incidence and mortality.
	
	The primary aim of this study is to estimate rates in rare cancers. To do so, we develop a new multivariate spatio-temporal approach that combines ideas derived from the above mentioned shared component models and the spatio-temporal interactions defined by \cite{knorr2000}. Based on this methodology, we will focus on the analysis of incidence and mortality rates for both pancreatic cancer and leukemia among males in the 142 administrative healthcare districts across Great Britain (England, Scotland, and Wales) over 9 biennial periods (2002-2003, 2004-2005, ..., 2018-2019).
	
	The remaining sections of the paper are organized as follows. In Section~2, we provide an overview of the classical spatio-temporal shared-component models. We then proceed to describe new spatio-temporal models that incorporate flexible shared spatio-temporal effects and provide details on how to implement the new models using R-INLA. Identifiability issues are also carefully discussed. We assess model adequacy through a simulation study in Section~3. In Section~4 we analyze the real data using the new models. Finally, Section~5 concludes with a discussion.

	\section{Spatio-temporal models}\label{Section2:}

	\subsection{Specific interactions for each health outcome}
	
	To define the models, let $O_{itd}$, $n_{itd}$ and $r_{itd}$ be the observed number of cases, the population at risk and the rates in each area $i$, $i = 1, \dots, A$ at time $t$, $t = 1, \dots, T$ for $d=I$, incidence, and $d=M$, mortality. Then, conditional on the rates $r_{itd}$, we assume that the number of observed cases follows a Poisson distribution with mean $\mu_{itd}=n_{itd}r_{itd}$. That is,
	\begin{eqnarray*}
		O_{itI}|r_{itI}&\sim& Poisson(\mu_{itI}=n_{it}r_{itI}), \quad \quad
		\log \mu_{itI} = \log n_{it}+\log r_{itI}, \nonumber \\
		O_{itM}|r_{itM}&\sim& Poisson(\mu_{itM}=n_{it}r_{itM}), \quad \quad
		\log \mu_{itM} = \log n_{it}+\log r_{itM}. \nonumber
	\end{eqnarray*}
	To model the log rates, $\log r_{itd}$, we first consider a spatio-temporal multivariate model with a different intercept for each health outcome, a shared spatial component \citep{held2005}, a time effect specific for each health outcome and specific spatio-temporal interaction for each health outcome. Therefore, we assume that the log rates, $\log r_{itd}$, can be written as
	\begin{eqnarray}\label{Eq:1}
		\log r_{itI} &=&\alpha_I + \delta \kappa_{i} + \gamma_{tI} + \chi_{itI}, \nonumber \\
		\log r_{itM} &=&\alpha_M + \frac{1}{\delta} \kappa_{i} + \gamma_{tM} + \chi_{itM},
	\end{eqnarray}
	where $\alpha_d$ is a health outcome-specific intercept, $\delta$ is a scaling parameter, $\kappa_{i}$ represents the shared spatial component, $\gamma_{td}$ represents the time effect specific for each health outcome $d$ and $\chi_{itd}$ are the spatio-temporal interactions specific for each health outcome $d$. The spatio-temporal interaction has the same structure for incidence and mortality but the amount of smoothing can be the same or different.
	
	Considering our goal of developing a novel multivariate spatio-temporal approach that combines concepts from established shared component models and the spatio-temporal interactions defined by \cite{knorr2000}, we will now provide a more detailed description of the shared spatial component model referenced in \autoref{Eq:1}. As previously mentioned, the shared spatial component model represents a particular case within spatial factor modeling \citep{wang2003}. Shared component models constitute a simple technique to model several health outcomes. When employing these models, there is no requirement to empirically assess the dependency between the health outcomes under study; instead, it is presumed a priori. That is, the health outcomes being studied are related due to either similar shared spatial patterns or common risk factors. These models have been widely used to examine the spatial distribution of related diseases \citep{held2006, cramb2015, kazembe2015, retegui2021, law2020} or to analyze the spatial correlation among the incidence and mortality of the same diseases \citep{etxeberria2018, etxeberria2023, retegui2023}.  Specifically, in this study, we exploit the correlation between incidence and mortality rates among different rare cancer types. Moreover, in these models, the unknown scaling parameter $\delta > 0$ is included to accommodate varying risk gradients of the shared component for the two health outcomes. 

We assign the following priors to each parameter and effects defined in \autoref{Eq:1}
\begin{eqnarray*}
&&\alpha_d \sim N\left(0,1/0\textrm{.}001\right), \quad d=I,M \quad \quad \quad \quad \quad \quad  \delta \sim \textrm{Gamma}(10,10),\\
&&p(\boldsymbol{\kappa}) \propto \exp\left(\frac{-\tau_{\kappa}}{2} \boldsymbol{\kappa}^{'} \mathbf{R}_{\kappa} \boldsymbol{\kappa} \right), \quad \quad \quad \quad \quad \quad \quad \quad p(\mathbf{\boldsymbol{\gamma}_d}) \propto \exp\left(\frac{-\tau_{\gamma_d}}{2} \boldsymbol{\gamma}_d^{'} \mathbf{R}_{\gamma} \boldsymbol{\gamma}_d \right),\\
&&p(\boldsymbol{\chi}_{d}) \propto \exp\left(\frac{-\tau_{\chi_d}}{2} \boldsymbol{\chi}_{d}^{'} \mathbf{Q}_{\chi} \boldsymbol{\chi}_{d} \right)
\end{eqnarray*}
\noindent where $\mathbf{R}_{\kappa}, \mathbf{R}_{\gamma}$ and $\mathbf{Q}_{\chi}$ are structure matrices. Specifically, $\mathbf{R}_{\kappa}$	is the spatial neighbourhood structure matrix defined by adjacency, i.e., two areas are neighbours if they share a common border. In the $\mathbf{R}_{\kappa}$ matrix the $i$th diagonal element is equal to the number of neighbours of the $i$th geographical area. For $i\not =j$, $R^{\kappa}_{ij}=-1$ if $i$ and $j$ are neighbours and 0 otherwise. $\mathbf{R}_{\gamma}$ is determined by the temporal structure matrix of a first order random walk \citep[see ][p. 95]{rue2005} and the structure matrix $\mathbf{Q}_{\chi}$ represents any of the four spatio-temporal interaction types proposed by \cite{knorr2000}. In Type~I interactions all cells of the precision matrix are independent without any structure in space and time, that is $\mathbf{Q}_{\chi}= \mathbf{I}_{TA}$, where $\mathbf{I}_{TA}$ is the identity matrix of size $TA \times TA$. Type~II interactions consider a first order random walk for time with no structure in space, i.e. $\mathbf{Q}_{\chi}= \mathbf{R}_{\gamma} \otimes \mathbf{I}_A$. When the structure effect is defined in space and the unstructured effect in time we have Type~III interactions. In this case, $\mathbf{Q}_{\chi}= \mathbf{I}_T \otimes \mathbf{R}_{\kappa}$. Finally, when we have structure in time and space, we consider Type~IV interactions defined as $\mathbf{Q}_{\chi}= \mathbf{R}_{\gamma} \otimes \mathbf{R}_{\kappa}$. The structure matrices for the different type of interactions are summarized in \autoref{tab3.0}.
Additionally, to fit the models uniform vague priors have been defined for the precision parameters $\tau_{\kappa}$, $\tau_{\gamma_d}$ and $\tau_{\chi_d}$.

\begin{table}[t!]
\centering
\caption{Specification of the four types of space-time interaction proposed by \cite{knorr2000}. \label{tab3.0}}
\begin{tabular}{lcccc}
	\toprule
	Space-time interaction &\multicolumn{2}{c}{Structure in} && $\mathbf{Q}_{\chi}$ \\
	&Time & Space &&\\
	\midrule
	Type I &  &  && $\mathbf{I}_{T} \otimes \mathbf{I}_{A}$  \\
	\multicolumn{5}{c}{ }\\
	Type II & \checkmark &  &&  $\mathbf{R}_{\gamma} \otimes \mathbf{I}_A$ \\
	\multicolumn{5}{c}{ }\\
	Type III &  & \checkmark&& $\mathbf{I}_T \otimes \mathbf{R}_{\kappa}$ \\
	\multicolumn{5}{c}{ }\\
	Type IV&\checkmark & \checkmark&&$ \mathbf{R}_{\gamma} \otimes \mathbf{R}_{\kappa}$  \\
	\bottomrule
\end{tabular}
\end{table}

\subsection{Shared interactions with time-varying scaling parameters}

In the preceding section, we examine spatio-temporal models featuring specific spatio-temporal interactions for each health outcome. In this section, we introduce a model that incorporates shared spatio-temporal interactions among incidence and mortality, with the objective of enhancing the estimates for less prevalent cancer sites.

Our initial approach involves adopting the shared component model \citep{held2005} to define the spatio-temporal effect. Therefore, we maintain the shared component for the area and the time effect for each health outcome $d$ as in the previous section, but we add a shared component term for the spatio-temporal effect. Then, we assume that the log rates, $\log r_{itd}$, have the following decomposition
\begin{eqnarray}\label{Eq:2*}
\log r_{itI} &=&\alpha_I + \delta \kappa_{i} + \gamma_{tI} + \varrho \chi_{it}, \nonumber \\
\log r_{itM} &=&\alpha_M + \frac{1}{\delta} \kappa_{i} + \gamma_{tM} + \frac{1}{\varrho} \chi_{it}
\end{eqnarray}
where $\varrho$ is a scaling parameter and $\boldsymbol{\chi}$ is the shared spatio-temporal interaction with the following priors 		
\begin{equation*}
\varrho \sim \textrm{Gamma}(10,10)\quad \text{and} \quad p(\boldsymbol{\chi}) \propto \exp\left(\frac{-\tau_{\chi}}{2} \boldsymbol{\chi}^{'} \mathbf{Q}_{\chi} \boldsymbol{\chi} \right),
\end{equation*}

\noindent where $\mathbf{Q}_{\chi}$ has any of the structure matrices defined by \cite{knorr2000}.

This model could be restrictive, as we may need to modulate the spatio-temporal interactions between cancer incidence and mortality,  taking into account changes
in their relationship over time.
Hence, we propose a novel model that incorporates a time-varying scaling parameter in the spatio-temporal shared effect, thereby increasing the model's flexibility, i.e.
\begin{eqnarray}\label{Eq:2}
\log r_{itI} &=&\alpha_I + \delta \kappa_{i} + \gamma_{tI} + \varrho_{t} \chi_{it}, \nonumber \\
\log r_{itM} &=&\alpha_M + \frac{1}{\delta} \kappa_{i} + \gamma_{tM} + \frac{1}{\varrho_t} \chi_{it}.
\end{eqnarray}

\noindent By defining $\boldsymbol{r}_I = (r_{11I},r_{21I},\dots, r_{A1I},r_{12I}, \dots, r_{ATI})'$ and \newline $\boldsymbol{r}_M = (r_{11M},r_{21M},\dots, r_{A1M},r_{12M}, \dots, r_{ATM})'$, \autoref{Eq:2} can also be expressed in matrix form as
\begin{equation*}
\arraycolsep=1.2pt
\log \left(\begin{array}{c}
	\boldsymbol{r}_I \\
	\boldsymbol{r}_M
\end{array}\right) = \left(\begin{array}{cc}
	\boldsymbol{1}_{AT} & \boldsymbol{0}\\
	\boldsymbol{0} & \boldsymbol{1}_{AT}
\end{array}\right)\left(\begin{array}{c}
	\alpha_I \\
	\alpha_M
\end{array}\right) +
\left(\begin{array}{c}
	\delta\boldsymbol{Z_1}  \\
	\frac{1}{\delta}\boldsymbol{Z_1}
\end{array}\right){\kappa}
+
\left(\begin{array}{cc}
	\boldsymbol{Z_2} & \boldsymbol{0}\\
	\boldsymbol{0} & \boldsymbol{Z_2}
\end{array}\right)\left(\begin{array}{c}
	\boldsymbol{\gamma_I} \\
	\boldsymbol{\gamma_M}
\end{array}\right) +
\left(\begin{array}{c}
	\boldsymbol{Z_3}\\
	\boldsymbol{Z_3}^{-1}
\end{array}\right) \boldsymbol{\chi}
\end{equation*}
where $\boldsymbol{1}_{AT}$ is a column of ones of size $AT$,
$\boldsymbol{Z_1} = \textrm{col}_{1\leq k\leq T}(\boldsymbol{I}^k_{A})$,
$\boldsymbol{\kappa} = (\kappa_1,\kappa_2, \dots, \kappa_A)'$,
$\boldsymbol{Z_2} = \textrm{diag}_{1\leq k\leq T}(\boldsymbol{1}_A^k)$,
$\boldsymbol{\gamma_I} = (\gamma_{1I},\gamma_{2I}, \dots,
\gamma_{TI})'$, $\boldsymbol{\gamma_M} = (\gamma_{1M},\gamma_{2M},
\dots, \gamma_{TM})'$, $\boldsymbol{Z_3} =
\textrm{diag}\left(\boldsymbol{\varrho}\right) \otimes \mathbf{I}_A$,
$\boldsymbol{Z_3}^{-1}=
\textrm{diag}\left(\boldsymbol{\varrho}^{-1}\right) \otimes
\mathbf{I}_A$ and $\boldsymbol{\chi} = (\chi_{11},\chi_{21}, \dots,
\chi_{AT})'$.

The scaling parameters $\varrho_t$ are not necessarily required to be distinct for all time points $t$. To adapt the flexibility of the model, we define $l$ ($1\leq l\leq T$) as the suitable number of scaling parameters that can be adjusted based on the data under analysis, making the model more or less flexible as needed. If this is the case,
\begin{equation*}\label{Eq:3}
\boldsymbol{\varrho} = \left(\varrho_1\mathbf{1'}_{m_1},\varrho_2\mathbf{1'}_{m_2}, \dots, \varrho_l\mathbf{1'}_{m_l}\right)',
\end{equation*}
where $\varrho_k$ are scaling parameters, $m_k$ is the number of years with the same scaling parameter $\varrho_k$ where $\sum_{k\leq l}m_k=T$, and $\mathbf{1}_{m_k}$ are column of ones of size $m_k$.

We assume that the $\varrho_k$ scaling parameters are independent and, therefore, for each $\varrho_k$ scaling parameter we assign a $\textrm{Gamma}(10,10)$ prior. As in the previous model defined in Equation \autoref{Eq:1}, uniform vague prior have been assigned for the precision parameters $\tau_{\kappa}$, $\tau_{\gamma_d}$ and $\tau_{\chi}$.

A sensitivity analysis was conducted using uniform vague priors, log-gamma priors, and PC-priors for the precision and scaling parameters. The estimated rates showed no differences across these priors. However, disparities were observed in the estimated posterior distribution for $\tau_{\gamma_d}$ when using gamma priors compared to other priors. Further information regarding the implementation of this model can be found in Section~2.3.

\subsection{Model implementation in INLA} \label{Subsection:ModelFitting}
The integrated nested Laplace approximation (INLA) technique \citep{rue2009} is used to fit the models described above.
INLA can be implemented in the free software R through the R-package R-INLA \citep{martino2009,martino2014} (\url{www.r-inla.org}).

To fit shared component models, the \texttt{besag2} model is defined in R-INLA. However, the flexible shared spatio-temporal models presented here are not directly available in R-INLA. Therefore, we have implemented them using the \texttt{rgeneric} model to define this latent effect. This model allows the user to define latent model components in R.
An additional drawback arises as INLA does not allow to repeat a latent effect within the same model \citep{martins2013}, and in these particular models,  the spatio-temporal interaction is common to both health outcomes. Therefore, to implement such models, we rely on the \texttt{copy} feature defined in R-INLA \citep{martins2013}. This feature enables us to incorporate the same latent effect twice in our model, by generating an almost identical copy of the latent field that is required multiple times in the model formulation. More precisely, to define the flexible shared component models,  we denote the latent effect of the spatio-temporal component by
\begin{equation}
\mathbf{z} = \boldsymbol{Z_3}\boldsymbol{\chi}. \nonumber
\end{equation}
We then consider  an extended latent effect $\boldsymbol{x} = \left(\boldsymbol{z},\boldsymbol{z^*}\right)$ where $\boldsymbol{z^*}$ is the almost identical copy of $\boldsymbol{z}$. Moreover, it is also possible for the copied latent effect to have a scale parameter $\lambda$. Therefore, we define $\boldsymbol{z^*}$ as
\begin{equation*}
\boldsymbol{z^*} = \lambda \boldsymbol{z} + \boldsymbol{\epsilon}
\end{equation*}
where $\boldsymbol{\epsilon}$ is a tiny error that controls the degree of closeness between $\boldsymbol{z}$ and $\boldsymbol{z}^*$. In our context, we need to consider that the copied latent effect defined in the model is $\boldsymbol{Z_3}^{-1}\boldsymbol{\chi}$. As such, it is necessary that $\boldsymbol{Z_3}^{-1}\boldsymbol{\chi} = \lambda \boldsymbol{z}$, which implies that the value of the unknown scale parameter must be $\lambda = \left(\boldsymbol{Z_3}^{-1}\right)^2$. Therefore to implement the flexible shared component model we have defined the copied latent effect $\boldsymbol{z^*}$ as
\begin{eqnarray}
\mathbf{z^*} &=& \boldsymbol{Z_3}^{-1}\boldsymbol{\chi} + \boldsymbol{\epsilon} = \left(\boldsymbol{Z_3}^{-1}\right)^2\boldsymbol{z} + \boldsymbol{\epsilon}. \label{Eq:5} \nonumber
\end{eqnarray}
Additionally, the structure of $\boldsymbol{z}^*$ is inherited from $\boldsymbol{\epsilon}$ and hence we define a spatio-temporal structure for $\boldsymbol{\epsilon}$. Therefore, $\boldsymbol{\epsilon}$ follows a Gaussian distribution with mean $\boldsymbol{0}$ and precision matrix $\tau_{\epsilon}\mathbf{Q}_{\chi}$. To achieve an almost identical copy of $\boldsymbol{z}$, we set a high precision value, specifically, $\tau_{\epsilon} = \exp(15)$  following the approach by \cite{martins2013}.

To implement the extended latent effect $\boldsymbol{x}$ with the \texttt{rgeneric} model, we need to define the distribution of $\boldsymbol{x}$, i.e.,
\begin{equation}\label{Eq:7}
\pi\left(\boldsymbol{x}\right) = \pi\left(\boldsymbol{z}\right)\pi\left(\boldsymbol{z}^*|\boldsymbol{z}\right). \nonumber
\end{equation}
After some algebra (see \autoref{AppendixA}), we obtain that $\boldsymbol{x}$ is distributed as
\begin{equation}
\boldsymbol{x} \sim N\left(\boldsymbol{0},\boldsymbol{Q_x}\right) \nonumber
\end{equation}
where the precision matrix $\boldsymbol{Q_x}$ is given by
\begin{equation}
\boldsymbol{Q_x} = \left(\begin{array}{cc}
	\tau_{\chi}\boldsymbol{Z_3}^{-1}\boldsymbol{Q}_\chi\boldsymbol{Z_3}^{-1}+\tau_{\epsilon}\left(\boldsymbol{Z_3}^{-1}\right)^2\boldsymbol{Q}_\chi\left(\boldsymbol{Z_3}^{-1}\right)^2& -\tau_{\epsilon}\left(\boldsymbol{Z_3}^{-1}\right)^2\boldsymbol{Q}_\chi\\
	-\tau_{\epsilon}\boldsymbol{Q}_\chi\left(\boldsymbol{Z_3}^{-1}\right)^2 & \tau_{\epsilon}\boldsymbol{Q}_\chi
\end{array}\right). \nonumber
\end{equation}

Once we have determined the distribution of the latent effect that needs to be fitted using the \texttt{rgeneric} function, we can proceed to define the required model using the \texttt{inla.rgneneric.define()} function. Finally, to fit the model, the procedure remains the same as for any other model readily available in R-INLA, using the \texttt{f()} function. The code used to implement all the models will be available at \url{https://github.com/spatialstatisticsupna/Shared_interactions}.

\subsubsection{Identifiability issues}

The proposed models incorporate health outcome-specific intercepts, a shared component model for space, a first order random walk for time, and flexible shared spatio-temporal effects.
As the spatial and  temporal random effects implicitly include an intercept, identifiability issues arise and constraints are needed. A quick solution is to put sum to zero constraints on the spatial and temporal effects. The interaction effects also overlap with the main spatial and temporal terms needing the inclusion of additional constraints \citep{goicoa2018}. In particular, for  Type~I and Type~IV interactions, the ones selected in the real data analyses below, the required constraints are $\sum_{i=1}^{A} \sum_{t=1}^{T} \chi_{it} = 0$ (Type~I ), and $\sum_{i=1}^{A} \chi_{it} = 0, \forall t$ and $\sum_{t=1}^{T} \chi_{it} = 0, \forall i$ (Type~IV).

In addition, shared component models have identifiability issues with the scaling parameter. According to \cite{held2005}, when generalizing the shared component model to two or more scaling parameters, it becomes necessary to impose the constraint $\sum_{k=1}^{n_k} \log \delta_k = 0$ where $n_k$ denotes the number of health outcomes under analysis and $\delta_k$ the corresponding scaling parameters. This constraint is automatically fulfilled in the shared component model for two health outcomes because for two scaling parameters, the sum to zero constraint on the logarithmic scale simply translates to $\delta_2 = 1/\delta_1$.
In the extension of the shared component model with time-varying scale parameter, we have $n_k=2*l$ scaling parameters since we have two health outcomes under analysis and $l$ scaling parameters for each health outcome. Therefore, it is necessary to satisfy the constraint $\sum_{k=1}^{2*l} \log \delta_k = 0$. Considering the definition of the scaling parameters in the flexible shared component model as $\delta_k = \varrho_k $ for incidence and $\delta_k = 1/\varrho_k$ for mortality, the sum to zero constraint is automatically satisfied, i.e.,

\begin{eqnarray}
\sum_{k=1}^{2*l} \log \delta_k &=& \sum_{k=1}^{l} \log \delta_k + \sum_{k=l+1}^{2*l} \log \delta_k = \sum_{k=1}^{l} \log \varrho_k + \sum_{k=1}^{l} \log \frac{1}{\varrho_k} \nonumber \\
&=& \sum_{k=1}^{l} \left(\log \varrho_k + \log \frac{1}{\varrho_k}\right) = \sum_{k=1}^{l} \log \left(\varrho_k \frac{1}{\varrho_k}\right) = \sum_{k=1}^{l} \log 1 = 0. \nonumber
\end{eqnarray}

\section{Simulation Study}\label{Section3:SimulationStudy}
To evaluate the performance of the new spatio-temporal models with flexible shared components, we conducted simulation studies. These studies are based on the spatial and temporal layout of the 142 health areas and 9 study periods of Great Britain, the one used in the real data analysis of Section~4. Additionally, we conduct a simulation study under the models defined in Section~2. For this reason, we define three different scenarios to simulate the log rates, $\log r_{itI}$ and $\log r_{itM}$, according to the spatio-temporal effects proposed previously. More precisely, the log rates are simulated as follows:

\begin{table}[t!]
\caption{\label{tab1} True values of parameters in Scenarios 1 to 3.}
\begin{tabular}{lcccccccccccc}
	\toprule
	Scenario & $\alpha_I$ & $\alpha_M$ & $\delta$ & $\tau_{\kappa}$ & $\tau_{\gamma_I}$ & $\tau_{\gamma_M}$ & $l$ & $\varrho_{k}$ & $m_{k}$ & $\tau_{\chi}$ & $\tau_{\chi_I}$ & $\tau_{\chi_M}$\\
	\midrule
	1 &  -8.9 & -9.3 & 0.9 & 35.7 & 200 & 120 & &&& & 400 & 550\\
	\multicolumn{13}{c}{ }\\
	2 & -8.9 & -9.3 & 0.9 & 35.7 & 200 & 120 & 1 & $\varrho_1 = 1.4$ & $m_1 = T$& 80 & &\\
	\multicolumn{13}{c}{ }\\
	\multirow{3}{*}{3} &\multirow{3}{*}{-8.9} & \multirow{3}{*}{-9.3}& \multirow{3}{*}{0.9}& \multirow{3}{*}{35.7}& \multirow{3}{*}{200}& \multirow{3}{*}{120} & \multirow{3}{*}{3} & $\varrho_1 = 1.0$ & $m_1 =3$ & \multirow{3}{*}{80} & & \\
	&&&&&&&& $\varrho_2 = 1.4$ & $m_2 =3$ &&&\\
	&&&&&&&& $\varrho_3 = 1.8$ & $m_3 =3$ &&&\\
	\bottomrule
\end{tabular}
\end{table}

\begin{itemize}
\item \textbf{Scenario 1:} Rates are generated using the  spatio-temporal interactions specific for each health outcome given by \autoref{Eq:1}. That is, we generate two spatio-temporal interactions, namely $\boldsymbol{\chi_I}=\left(\chi^I_{11}, \chi^I_{21}, \dots, \chi^I_{AT}\right)$ and $\boldsymbol{\chi_M}=\left(\chi^M_{11}, \chi^M_{21}, \dots, \chi^M_{AT}\right)$ where $p(\boldsymbol{\chi}_{d}) \propto \exp\left(\frac{-\tau_{\chi_d}}{2} \boldsymbol{\chi}_{d}^{'} \mathbf{Q}_{\chi_d} \boldsymbol{\chi}_{d} \right)$ ($d=I,M$). The true values assume for the variance components $\tau_{\chi_d}$ for $d=I,M$ can be found in \autoref{tab1}.

\item \textbf{Scenario 2:} Rates are generated using the shared spatio-temporal effects given by \autoref{Eq:2*}. That is, $\boldsymbol{\zeta} = \left(\varrho\boldsymbol{\chi},\frac{1}{\varrho}\boldsymbol{\chi}\right)$ where  $p(\boldsymbol{\chi}) \propto \exp\left(\frac{-\tau_{\chi}}{2} \boldsymbol{\chi}^{'} \mathbf{Q}_{\chi} \boldsymbol{\chi} \right)$. Note that this scenario can be rewritten as a special case of the flexible shared spatio-temporal effect given by \autoref{Eq:2} with a unique scaling parameter for all time periods.

\item  \textbf{Scenario 3:} Rates are generated using the flexible shared spatio-temporal effects given by \autoref{Eq:2}. That is, $\boldsymbol{\zeta} = \left(\boldsymbol{Z_3}\boldsymbol{\chi}, \boldsymbol{Z_3}^{-1}\boldsymbol{\chi}\right)$ where $\boldsymbol{Z_3} =
\textrm{diag}\left(\boldsymbol{\varrho}\right) \otimes \mathbf{I}_A$ and $p(\boldsymbol{\chi}) \propto \exp\left(\frac{-\tau_{\chi}}{2} \boldsymbol{\chi}^{'} \mathbf{Q}_{\chi} \boldsymbol{\chi} \right)$. We define three scaling parameters for each three time periods, that is $l=3$ and $m_1=m_2=m_3=3$.
\end{itemize}

In all scenarios we maintain the same intercepts and spatial and temporal effects. Precisely, the $\alpha_j$ ($j=I,M$) and $\delta$ are fixed constants and $\boldsymbol{\kappa}$ and $\boldsymbol{\gamma_d}$ are generated from the models proposed in Section~2; specifically $p(\boldsymbol{\kappa}) \propto \exp\left(\frac{-\tau_{\kappa}}{2} \boldsymbol{\kappa}^{'} \mathbf{R}_{\kappa} \boldsymbol{\kappa} \right)$ and $p(\mathbf{\boldsymbol{\gamma}_d}) \propto \exp\left(\frac{-\tau_{\gamma_d}}{2} \boldsymbol{\gamma}_d^{'} \mathbf{R}_{\gamma} \boldsymbol{\gamma}_d \right)$ where the spatial neighbourhood matrix $\mathbf{R}_{\kappa}$ is based on the Great Britain map and $\mathbf{R}_{\gamma}$ is determined by the temporal structure of a first order random walk. The structure matrices $\mathbf{Q}_{\chi_d}$, in Scenario 1, and $\mathbf{Q}_{\chi}$, in Scenarios 2 and 3, have any of the structures defined by \cite{knorr2000}. Therefore, each of our scenarios will have four sub-scenarios, one for each structure matrix. The true values of the parameters assumed by each Scenario are shown in \autoref{tab1}.

To assess the performance of our proposed models, we simulated $N=100$ data sets for each sub-scenario by assuming the data $O_{itd}$ arise from a Poisson model
\begin{eqnarray*}
O_{itI}|r_{itI}&\sim& Poisson(\mu_{itI}=n_{it}r_{itI}),  \nonumber \\
O_{itM}|r_{itM}&\sim& Poisson(\mu_{itM}=n_{it}r_{itM}), \nonumber
\end{eqnarray*}
where $n_{it}$ are the population of the real data analysis of Section~3. We fitted three different models to every scenario. Specifically, we fitted models described by \autoref{Eq:1}, denoted as Model 1; \autoref{Eq:2*}, named Model 2; and Model 3, which corresponds with \autoref{Eq:2} with $l=3$ and $m_1=m_2=m_3=3$.

To compare the models we compute the differences in Deviance Information criterion \citep[DIC,][]{spiegelhalter2002}, Watanabe-Akaike Information criterion \citep[WAIC,][]{watanabe2010} and logarithmic score \citep[LS,][]{gneiting2007} between each model and the true model for each Scenario, i.e. $DIC_k - DIC_{true}$ for models $k = 1, 2, 3$ and similarly for other scores. Consequently, negative values indicate superiority over the true model.

Moreover, as predictive measures we calculate the mean absolute relative bias (MARB) and the mean relative root mean square error (MRRMSE); and for a proper scoring rule we employ the interval score \citep[IS,][]{gneiting2007},
\begin{eqnarray*}
MARB &=& \left(\frac{1}{NATD}\sum_{j=1}^{N}\sum_{i=1}^{A}\sum_{t=1}^T\sum_{d=1}^{D} \frac{|r^j_{itd} - r_{itd}|}{r_{itd}}\right)*100\\
MRRMSE &=& \left(\frac{1}{N}\sum_{j=1}^{N}\sqrt{\frac{1}{ATD}\sum_{i=1}^{A}\sum_{t=1}^T\sum_{d=1}^{D} \left(\frac{r^j_{itd} - r_{itd}}{r_{itd}}\right)^2}\right)*100\\
IS &=& \frac{1}{NATD}\sum_{j=1}^{N}\sum_{i=1}^{A}\sum_{t=1}^T\sum_{d=1}^{D}\left( (u-l)+\frac{2}{\beta}(l-r_{itd})\boldsymbol{1}\left\{r_{itd}<l\right\}\right. \\
&& \left. + \frac{2}{\beta}(r_{itd}-u)\boldsymbol{1}\left\{r_{itd}>u\right\}\right)\\
\end{eqnarray*}
where $j$ is the simulation number, $i$ is the area, $t$ is the year, $d$ represents the health outcome ($d= I,M$), $r^j_{itd}$ is the estimated rate in simulation $j$ for area $i$, time $t$ and health outcome $d$, $r_{itd}$ is the true value in the simulation study, $l=q^j_{itd;\beta/2}$ and $u=q^j_{itd;1-\beta/2}$ are the $\beta/2$ and $1-\beta/2$ quantiles of the posterior distribution of the fitted incidence rate for simulation $j$, area $i$, time $t$ and health outcome $d$, and $\boldsymbol{1}\left\{.\right\}$ is the indicator function that takes value 1 if the event in brackets is true and 0 otherwise.

We also calculate the percentage change in MARB, MRRMSE and IS for each model compared to the true model in each scenario, i.e. $\Delta^{MARB}_k= (MARB_k - MARB_{true})/MARB_{true}*100$ and similarly for other measures. Additionally, we compute the credible interval length, denoted as $CIL = u-l$, and the coverage percentage for $\beta=0.05$.

\subsection{Results}

\begin{table}[pt!]
\centering
\caption{Percentiles of DIC and LS difference between the true model and the other models.  Symbol -  indicates that no differences are provided since it represents the true model.} \label{tab1*}
\resizebox{0.83\textwidth}{!}{
	\begin{tabular}{lrrrcrrrcrrr}
		\toprule
		&\multicolumn{3}{c}{Scenario 1} & & \multicolumn{3}{c}{Scenario 2} & & \multicolumn{3}{c}{Scenario 3}\\
		\midrule
		& \multicolumn{3}{c}{DIC} &&	\multicolumn{3}{c}{DIC}&& \multicolumn{3}{c}{DIC} \\
		\midrule
		& \%2.5 & \%50 & \%97.5  && \%2.5 & \%50 & \%97.5  && \%2.5 & \%50 & \%97.5\\
		&\multicolumn{11}{l}{Type I}\\		
		Model 1 & - & -  & - &  & 139.23 & 188.18 & 246.69 && 159.03 & 223.00 & 283.20  \\
		Model 2 & -1.17 & 18.14 & 40.65 &  &  - & -  & -  && 34.44 & 71.27 & 101.43 \\
		Model 3 & -2.42 & 21.59 & 46.18 & &  -6.09 & 2.45 & 8.03  &&  - & -  & -  \\
		\multicolumn{12}{l}{ }\\	
		&\multicolumn{11}{l}{Type II}\\	
		Model 1 &  - & -  & -  && 126.71 & 176.76 & 215.89  && 148.53 & 199.04 & 253.26 \\
		Model 2 &  15.22 & 47.85 & 92.82  &&  - & -  & -  && 48.78 & 75.13 & 119.4 \\
		Model 3 & 16.53 & 52.65 & 100.79 &&  -7.04 & 2.80 & 7.46  && - & -  & -   \\
		\multicolumn{12}{l}{ }\\	
		&\multicolumn{11}{l}{Type III}\\	
		Model 1 &  - & -  & -  && 33.86 & 67.59 & 105.87   && 53.21 & 87.98 & 128.19  \\
		Model 2 &  -3.78 & 6.82 & 30.72  &&  - & -  & -  && 13.48 & 33.64 & 62.16\\
		Model 3 & -4.68 & 7.64 & 25.83  && -7.86 & 2.46 & 7.12  &&  - & -  & -  \\
		\multicolumn{12}{l}{ }\\	
		&\multicolumn{11}{l}{Type IV}\\	
		Model 1 &   - & -  & -  && 43.85 & 81.44 & 107.80   && 59.30 & 95.23 & 137.72  \\
		Model 2 & 4.93 & 21.07 & 44.76 &&  - & -  & -   && 16.61 & 38.93 & 62.72  \\
		Model 3 & 8.41 & 27.49 & 51.71  &&  -7.42 & 1.97 & 7.55  && - & -  & - \\
		\bottomrule
		
		\midrule
		&  \multicolumn{3}{c}{LS} && \multicolumn{3}{c}{LS} &&  \multicolumn{3}{c}{LS}\\
		\midrule
		&  \%2.5 & \%50 & \%97.5 && \%2.5 & \%50 & \%97.5&& \%2.5 & \%50 & \%97.5  \\
		&\multicolumn{11}{l}{Type I}\\		
		Model 1 &  -  & - & -  &  &   175.80 & 221.18 & 267.79 &&  178.55 & 226.62 & 270.90 \\
		Model 2 &  -0.56 & 8.24 & 17.62 &  &  -  & - & -&& 7.50 & 23.18 & 36.96  \\
		Model 3 & -0.86 & 10.31 & 21.17& &   -4.17 & 2.36 & 6.38 &&  - & -  & -   \\
		\multicolumn{12}{l}{ }\\	
		&\multicolumn{11}{l}{Type II}\\	
		Model 1 &  -  & - & - && 114.21 & 146.26 & 176.04 &&  144.99 & 184.66 & 228.67\\
		Model 2 &  8.07 & 24.98 & 47.37  &&  -  & - & -&& 17.17 & 34.19 & 55.23 \\
		Model 3 &  10.14 & 30.42 & 54.70  &&  -4.80 & 1.74 & 5.77 && - & -  & -  \\
		\multicolumn{12}{l}{ }\\	
		&\multicolumn{11}{l}{Type III}\\	
		Model 1 &  -  & - & -  &&  29.74 & 49.06 & 73.85 && 35.80 & 58.48 & 86.60 \\
		Model 2 &   -1.36 & 3.58 & 14.54  &&  -  & - & - &&  6.84 & 17.36 & 31.50 \\
		Model 3 &  -1.30 & 4.63 & 13.66 && -4.04 & 1.28 & 3.29 &&  - & -  & -  \\
		\multicolumn{12}{l}{ }\\	
		&\multicolumn{11}{l}{Type IV}\\	
		Model 1 &  -  & - & -  && 34.35 & 53.67 & 72.74 &&  40.26 & 64.8 & 94.32 \\
		Model 2 &  2.70 & 11.13 & 23.91 && -  & - & -  &&  9.92 & 21.30 & 37.39 \\
		Model 3 &   5.61 & 15.78 & 27.55  &&  -5.03 & 1.20 & 4.75 && - & -  & -  \\
		\bottomrule
	\end{tabular}
	}
\end{table}

\autoref{tab1*} summarizes the difference in DIC and LS providing the $2.5$, $50$ and $97.5$ percentiles of the difference. In Scenario 1, Model 2 and Model 3 perform as well as the true models in terms of both difference in DIC and LS with larger disparities in Type II interaction. 
In scenarios 2 and 3, the true models easily beats Model 1. As expected, in Scenario 2, Model 2 and Model 3 perform similarly, indicating that Model 3 can effectively estimate all the different scaling parameters with similar values when the data requires. Finally, in Scenario 3, the true model also beats Model 2 but the disparities are narrower than those observed against Model 1.
Additionally, less disparities among models are seen for Type III interaction. This could be attributed to the fact that Type III interaction has structure in space and not in time, and Model 3 is more flexible in terms of time than space due to the time-varying scale parameter. The largest disparities among the true models and the other models are observed for Type II interactions, except for Model 3 in Scenario 2 as previously mentioned. 
Result for WAIC can be found in \autoref{AppendixB} and align with the same conclusions reached using DIC and LS.

\begin{table}[t!]
\centering
\caption{MARB and IS values, together with the percentage change for each measure relative to the true model in each simulation study. Symbol - indicates that no differences are provided since it represents the true model.} \label{tab1**}
\resizebox{\textwidth}{!}{
	\begin{tabular}{lrrrrcrrrrcrrrr}
		\toprule
		&\multicolumn{4}{c}{Scenario 1} & & \multicolumn{4}{c}{Scenario 2} & & \multicolumn{4}{c}{Scenario 3}\\
		\midrule
		& MARB & $\Delta^{MARB}$ & IS & $\Delta^{IS}$  && MARB & $\Delta^{MARB}$ & IS & $\Delta^{IS}$ && MARB & $\Delta^{MARB}$ &  IS & $\Delta^{IS}$  \\
		\midrule
		&\multicolumn{14}{l}{Type I}\\		
		Model 1 & 4.86 & -  & 3.35 & -  && 7.87 & 11.32  & 5.43 & 9.91 && 8.03 & 12.94 &  5.72 & 14.31 \\
		Model 2 & 4.92 & 1.23 & 3.74 & 11.55 &&  7.07 & -  &  4.94 & -&& 7.38 & 3.80  & 5.43 & 8.64  \\
		Model 3 & 4.93 & 1.44 &  3.69 & 10.04 && 7.07 & 0.00 &  4.95 & 0.22 && 7.11 &  - & 5.00 & -   \\
		\multicolumn{15}{l}{ }\\	
		&\multicolumn{14}{l}{Type II}\\	
		Model 1 &  5.05 &  -  & 3.46 & -  && 7.58 & 13.98  & 5.23 & 11.85  &&   8.05 & 13.70 &  5.78 & 13.96 \\
		Model 2 & 5.24 & 3.76 &  4.10 & 18.51 &&   6.65 &  - &  4.67 & -   &&  7.36 & 3.95 &  5.55 & 9.57 \\
		Model 3 & 5.26 & 4.16 & 4.01 & 15.76  && 6.66 & 0.15 &  4.68 & 0.26 &&  7.08 & -  & 5.07 & - \\
		\multicolumn{15}{l}{ }\\	
		&\multicolumn{14}{l}{Type III}\\	
		Model 1 &  4.13 & - &  2.77 & -   && 5.96 & 7.19 &  4.16 & 6.17 && 6.06 & 8.21 & 4.36 & 10.71\\
		Model 2 & 4.19 & 1.45 & 2.90 & 4.81  &&  5.56 & -  &  3.92 &  -  && 5.77 & 3.04 &  4.23 & 7.52  \\
		Model 3 & 4.20 & 1.69 &  2.90 & 4.80  &&  5.58 & 0.36 &  3.94 & 0.45  && 5.60 & -  &  3.93 & -  \\
		\multicolumn{15}{l}{ }\\	
		&\multicolumn{14}{l}{Type IV}\\	
		Model 1 & 4.27 & -  &  2.88 & -   && 5.98 & 9.32 &  4.17 & 7.90 && 6.28 & 9.41 &  4.56 & 12.26\\
		Model 2 & 4.42 & 3.51 & 3.16 & 9.92 && 5.47 & -  &  3.86 &  -  && 5.93 & 3.31 & 4.38 & 7.75\\
		Model 3 & 4.45 & 4.22 &  3.17 & 10.04 && 5.48 & 0.18 &  3.87 & 0.22 &&  5.74 & - & 4.07 & - \\
	\bottomrule
	\end{tabular}
	}
\end{table}

\autoref{tab1**} presents the MARB and IS values and the percentage change for each of the measures. The results observed for MARB are reasonably consistent with those presented in  \autoref{tab1*}, in terms that large DIC differences generally corresponds to large $\Delta^{MARB}$. Regarding IS results, consistency can be observed in Scenario 2 and Scenario 3 with the results observed in  \autoref{tab1*}. However, disparities are noticeable in Scenario 1. The worst results obtained by Model 2 and Model 3 can be attributed to the lower coverage percentage achieved by these models, as they have smoother credible intervals (see \autoref{AppendixB}).
Results for MRRMSE, 95\% credible interval lengths and coverage percentages can be found in \autoref{AppendixB}. We can see that results for MRRMSE align with what is observed for MARB. Model 2 and Model 3 present slimmer credible intervals than Model 1 but the coverage is around 95\%, except for Scenario 1 as we have mentioned.

In terms of computational cost, all scenarios show similar execution times. \autoref{tab3.1} shows the average computational cost for each model based on the interaction type defined. As expected, the computational cost increases with the complexity of the interaction type. Notably, Model 3, which incorporates flexible shared spatio-temporal terms, has the highest computational cost, ranging from 81 seconds with Type I interaction to 30 minutes with Type IV interaction. Computations were performed on a computer with a 3.00 GHz Intel Core i5-9500 processor and 20GB RAM, using the stable version of R-INLA INLA\_22.12.16.

\begin{table}[b!]
\centering
\caption{\label{tab3.1} CPU running time in seconds, as provided by the inla function.}
\begin{tabular}{lcccc}
	\toprule
	& Type I & Type II & Type III & Type IV\\
	\midrule
	Model 1 & 4.34 & 14.12 & 17.45 & 39.08 \\
	\multicolumn{5}{c}{ }\\
	Model 2 & 45.51 & 58.73 & 200.85 & 1004.03\\
	\multicolumn{5}{c}{ }\\
	Model 3 & 80.64 & 128.94 & 262.84 & 1739.65\\
	\bottomrule
\end{tabular}
\end{table}

\section{Real Data Analyses}\label{Section3:Illustration}

\begin{figure}[b!]
	\begin{center}
		\scalebox{0.3}{\includegraphics[page=1]{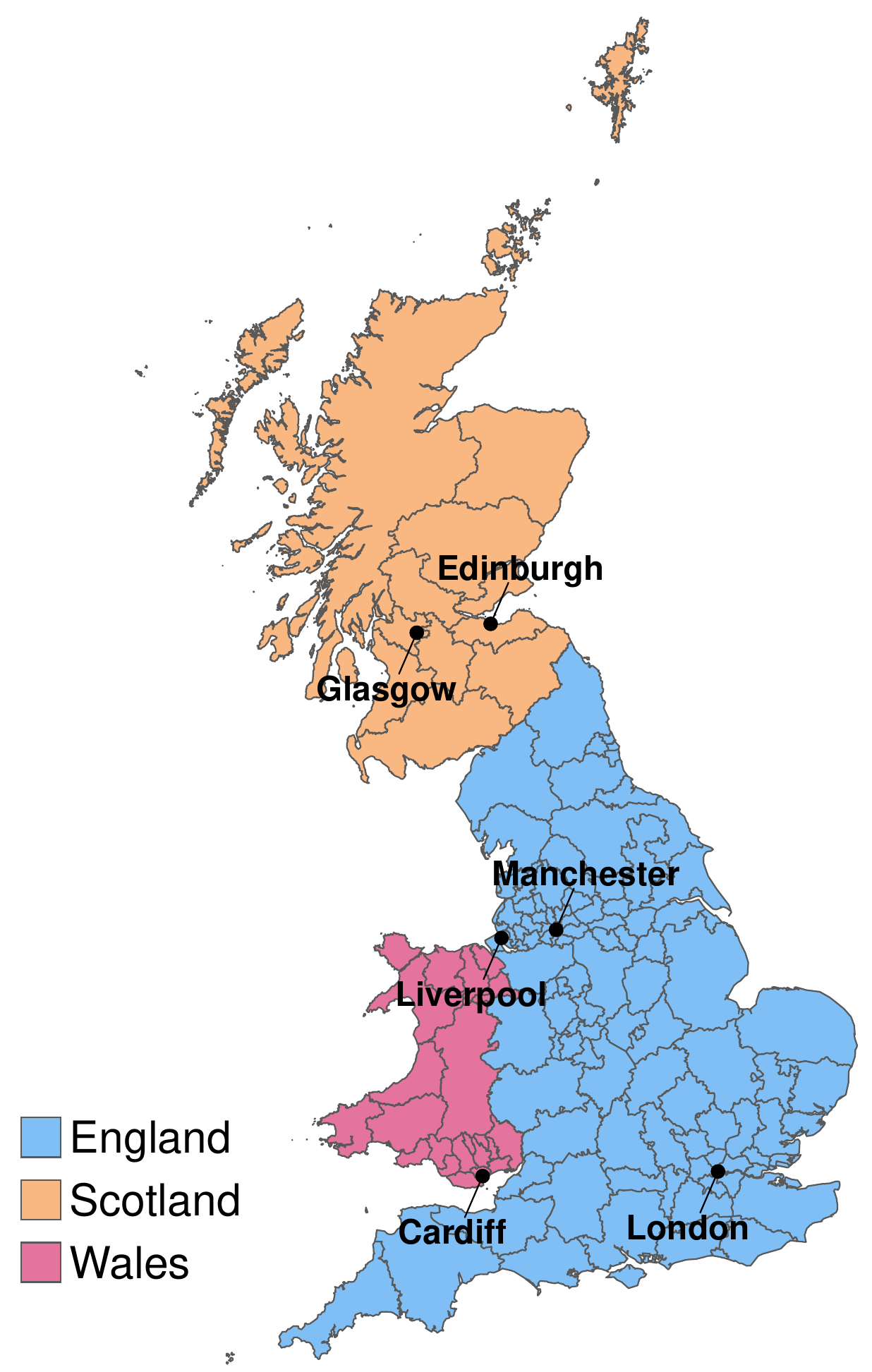}}
	\end{center}
	\caption{\label{fig1} Administrative division of Great Britain: 106 levels of English clinical commissioning groups (blue), 22 levels of Welsh local authorities (pink) and 14 levels of Scottish health boards (orange).}
\end{figure}

In this section, we use the novel spatio-temporal models with flexible shared spatio-temporal effects to analyze pancreatic cancer and leukaemia in British males.  These cancer sites belong to the rare cancer cohort and exhibit distinct characteristics. On one hand, pancreatic cancer is one of the most lethal cancers \citep{sung2021}, with survival rates in England remaining lower compared to similarly wealthy countries \citep{exarchakou2020}. High-risk factors for pancreatic cancer include smoking, alcohol consumption, and chronic pancreatitis. Recent studies have also highlighted the significance of blood type, glucose levels, and lipid levels in the development of pancreatic cancer \citep{zhao2020}. On the other hand, according to \cite[GLOBOCAN,][]{sung2021}, leukaemia was the 15th most commonly diagnosed cancer and the 11th leading cause of cancer mortality worldwide in 2020. The specific aetiology of leukaemia remains elusive, but some research suggests that these malignancies often develop in the context of genetic abnormalities, immunosuppression, and exposure to risk factors such as ionizing radiation or carcinogenic chemicals \citep{bispo2020}. Therefore, given the low incidence and mortality rates of both cancer types, spatial and/or temporal distributions have not been extensively studied in the literature. However, such an analysis could offer valuable insights into the disease distribution and the primary environmental or genetic risk factors that may influence it. Therefore, in this study, we propose a modeling approach to analyze the spatial and temporal distribution of these diseases. To begin, we present an exploratory data analysis.

\subsection{Exploratory data analysis}

The area under study corresponds to England, Wales and Scotland, which comprise the entire island of Great Britain, including small adjacent islands. The national health system of each region operates independently, thus the data have been collected separately and merged into a single database.
The URL for the original data sources of cancer and population data, along with the data used in this analysis, are available at \url{https://github.com/spatialstatisticsupna//Shared_interactions}.
Regarding the regions being examined, different territorial divisions exist. Here England has been divided at the clinical commissioning group level (106 regions), Wales at the local authority level (22 regions), and Scotland at the health board level (14 regions), resulting in a total of 142 small areas (see \autoref{fig1}).
The administrative division used for this study ranges from \mbox{9 544} to about \mbox{1 065 000} inhabitants per unit.

\begin{figure}[b!]
	\begin{center}
		\scalebox{0.047}{\includegraphics[page=1]{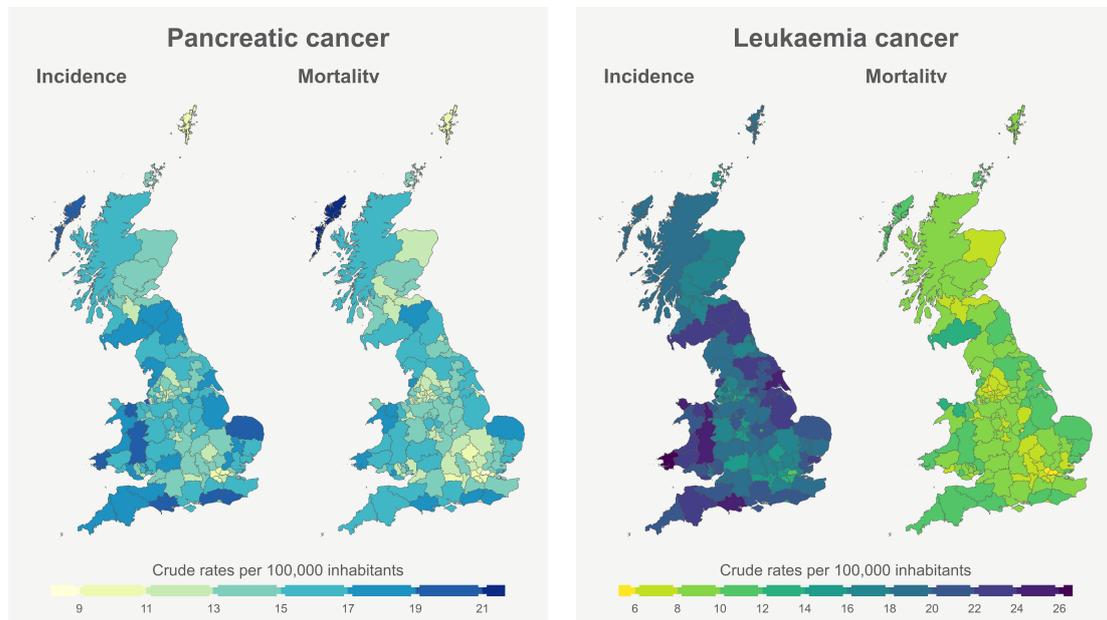}}
	\end{center}
	\caption{\label{fig2} Spatial distribution of crude incidence and mortality rates per \mbox{100 000} inhabitants for pancreatic cancer (left) and leukaemia (right) during the period 2002-2019.}
\end{figure}

During the 18 years of the study period (2002-2019), there were a total of \mbox{79 141} pancreatic cancer cases and \mbox{71 572} deaths, and \mbox{97 283} leukaemia cases and \mbox{45 768} deaths in Great Britain.  \autoref{fig2} shows the geographical patterns of crude incidence and mortality rates per \mbox{100 000} inhabitants for both cancer locations. The spatial patterns of pancreatic cancer incidence and mortality are very similar. The crude rates move between 8 and 22 cases or deaths per \mbox{100 000} inhabitants.
In general, the highest crude rates are found in the southern coastal areas of England and Wales, with the exception of the most westerly islands of Scotland, where very high rates are observed for both incidence and mortality. In contrast, areas in the central and northeastern parts of Great Britain have the lowest incidence and mortality rates.
On the contrary, leukaemia exhibits crude incidence rates ranging from 11 to 27 cases per \mbox{100 000} inhabitants, along with mortality rates ranging from 5 to 13 deaths per \mbox{100 000} inhabitants. \autoref{fig2} illustrates regional disparities in these rate values. Notably, elevated rates are observed in Wales, whereas Scotland generally exhibits lower rates, with the exception of the two regions bordering England, where higher rates are observed. In England, higher rates are observed in coastal areas, while the lowest rates are concentrated in two specific locations near London and Manchester.

\begin{figure}[b!]
	\begin{center}
		\scalebox{0.39}{\includegraphics[page=1]{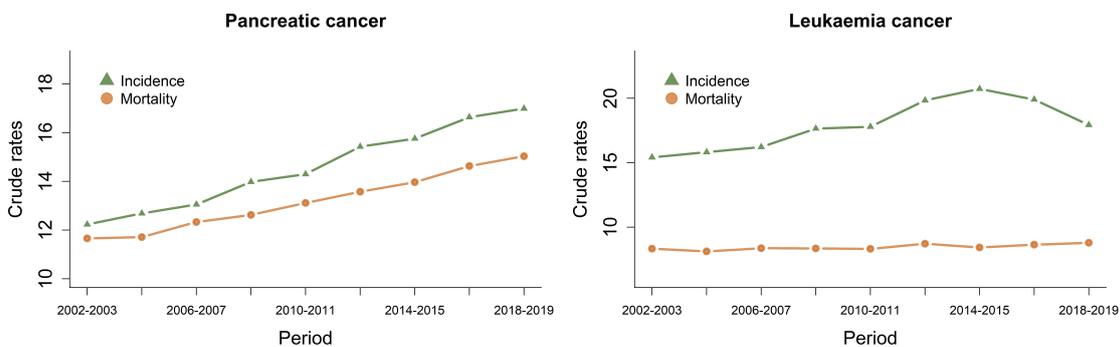}}
	\end{center}
\vspace*{-0.7cm}
	\caption{\label{fig3} Temporal trend of crude incidence and mortality rates per \mbox{100 000} inhabitants throughout the study period (2002-2019) for pancreatic cancer (left) and leukaemia (right) computed in biennial periods.}
\end{figure}

\autoref{fig3} displays the temporal trend of crude incidence and mortality rates per \mbox{100 000} inhabitants during the study period, computed in biennial intervals. We observe that for pancreatic cancer, both incidence and mortality rates have shown an upward trend over time. Conversely, in the case of leukaemia, we observe a predominantly linear growth in incidence until 2010-2011, followed by a pattern resembling an inverted U-shape. However, the mortality rate for leukaemia remains relatively stable throughout the study period.

\begin{figure}[t!]
\begin{center}
	\scalebox{0.31}{\includegraphics{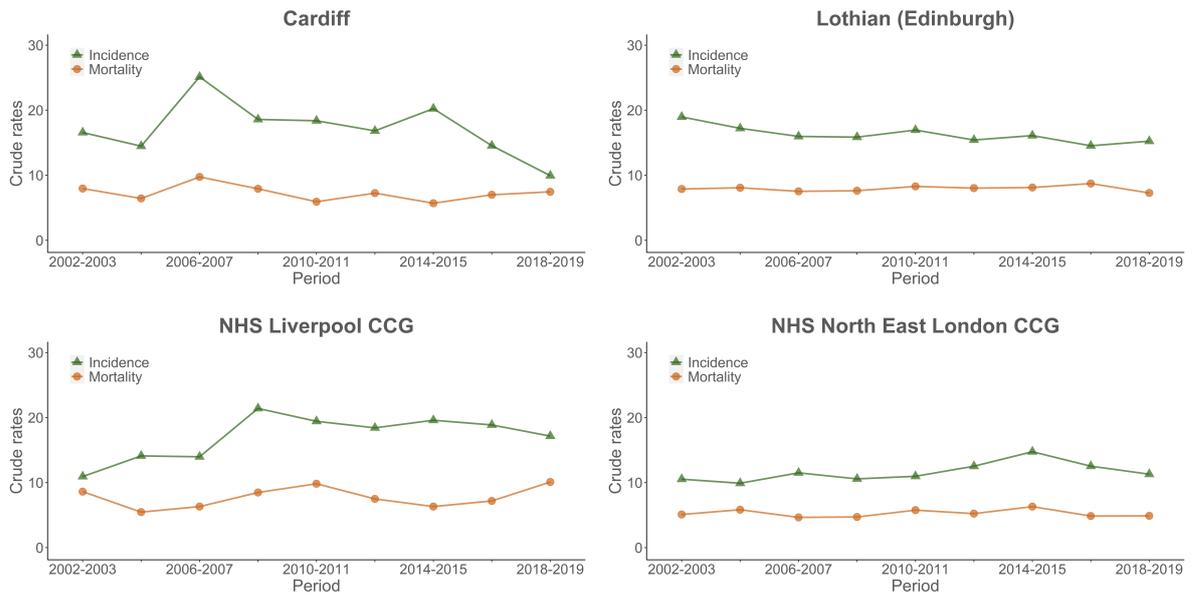}}
\end{center}
\caption{\label{fig3.2} Evolution of crude area-specific incidence and mortality rates per \mbox{100 000} inhabitants throughout the study period for pancreatic cancer (top) and leukaemia (bottom) in Cardiff, Lothian (Edinburgh), NHS Liverpool CCG and NHS North East London CCG.}
\end{figure}

\autoref{fig3.2} shows the temporal trends in incidence and mortality rates for four specific areas in Great Britain, namely Cardiff, Edinburgh, Liverpool, and North East London, for both cancer locations. Regarding pancreatic cancer, the incidence and mortality rates exhibit, in general, similar temporal trends for each area. Notably, there are discernible variations in the ratio between incidence and mortality rates within specific years (e.g. Cardiff year 2006-2007), as well as distinct trends observed in certain periods (e.g. Cardiff between the periods 2010-2011 and 2014-2015). A distinct pattern emerges when analyzing leukaemia, as we do not observe comparable trends across the selected areas. Instead, we notice that changes in trends occur during the same years for each area. For instance, in the case of Cardiff, both incidence and mortality exhibit similar trends until 2010-2011, but afterward, divergent trends become noticeable. We would like to remind that this is an exploratory data analysis; therefore, crude rates may exhibit significant variability, which needs to be smoothed using appropriate modeling techniques. In the next section, we model the rates to reach more reliable conclusions.

\subsection{Smoothing pancreatic and leukaemia rates using spatio-temporal models}

In this section, we explore various multivariate spatio-temporal models to conduct a comprehensive study of pancreatic cancer and leukaemia, focusing on the incidence and mortality rates among males in Great Britain from 2002 to 2019.
Prior to fitting multivariate spatio-temporal models, we fit a series of univariate spatio-temporal models considering different spatial, temporal, and spatio-temporal priors. In particular, we use both the intrinsic conditional autoregressive prior \citep[iCAR,][]{besag1991} and the scaled Besag-York-Molli{\'e} \citep{riebler2016} prior for the spatial random effect, a first- and second-order random walk for the temporal effects, and the
four types of interactions defined by \cite{knorr2000}. According to different model selection criteria DIC, 
WAIC 
and LS, 
for pancreatic cancer we finally select a model with iCAR prior for the spatial effect, first-order random walk for the temporal one and a Type~II interaction. For leukaemia, we select the same spatial and temporal priors but a Type~IV interaction (results not shown to save space). These selected models serve as the basis for defining several multivariate spatio-temporal models. In particular from now on, we only consider a random walk of first order for the temporal effect.

\begin{table}[b!]
\caption{\label{tab2} Fitted models and their performance in terms of the selection criteria for each cancer site.}
\resizebox{\textwidth}{!}{
	\begin{tabular}{cllrrrccrrr}
		\toprule
		&&&\multicolumn{3}{c}{Pancreatic cancer}&&&\multicolumn{3}{c}{Leukaemia cancer}\\
		\midrule
		&&& DIC & WAIC & LS &&& DIC & WAIC & LS\\
		\midrule
		\multicolumn{11}{l}{\bf Independent spatio-temporal interactions}\\
		&Model 1.1& Type II& 17070 & 17063 & 8545 && Type IV& 17084 & 17107 & 8601\\
		&Model 1.2& Type II& 17079 & 17080 & 8551 && Type IV& 17046 & 17054 & 8576\\
		&Model 1.3& Type II& 17083 & 17080 & 8552 && Type IV& 17050 & 17060 & 8579\\
		&Model 1.4& Type II& 17077 & 17069 & 8549 && Type IV& 17053 & 17061 & 8580\\
		\multicolumn{11}{l}{ }\\
		\multicolumn{11}{l}{\bf Shared spatio-temporal interactions}\\
		\multicolumn{11}{l}{\hspace*{0.2cm} $\boldsymbol{l=1}$}\\
		&\textbf{Model 3.1a}& \textbf{Type I}& \textbf{16716} & \textbf{16523} & \textbf{8289} && Type IV& 17032 & 17038 & 8567\\
		&Model 3.2a& Type I& 16720 & 16524 & 8290 && Type IV& 17002 & 16996 & 8545\\
		&Model 3.3a& Type I& 16723 & 16514 & 8287 && Type IV& 16999 & 16996 & 8545\\
		&Model 3.4a& Type I& 16727 & 16547 & 8300 && Type IV& 17003 & 16998 & 8547\\
		\multicolumn{5}{l}{\hspace*{0.2cm} $\boldsymbol{l=3}$}&& \multicolumn{5}{l}{\hspace*{0.2cm} $\boldsymbol{l=7}$}\\
		&Model 3.1b& Type I& 16719 & 16524 & 8290 && Type IV& 17043 & 17062 & 8573\\
		&\textbf{Model 3.2b}& Type I& 16718 & 16517 & 8287 && \textbf{Type IV}& \textbf{16984} & \textbf{16992} & \textbf{8538}\\
		&Model 3.3b& Type I& 16722 & 16527 & 8292 && Type IV& 17008 & 17012 & 8550\\
		&Model 3.4b& Type I& 16713 & 16508 & 8283 && Type IV& 17002 & 17007 & 8545\\
		\bottomrule
	\end{tabular}
}
\end{table}

We explore various sets of multivariate models, each encompassing different spatial effects and specific as well as shared spatio-temporal interactions for health outcomes.
While we implement all four types of interactions, we only include  in \autoref{tab2} the models with the interaction type that showed the best fit.
Models~1.1 to 1.4 have specific interactions for each health outcome  while  Models~3.1 to Model~3.4 have shared interactions. Model~1.1 is given by \autoref{Eq:1} and in Models 1.2 to 1.4 we  modify the spatial effect by adding different spatially unstructured random effects.  More precisely, in Model~1.2, we include a spatially unstructured random effect for mortality. In Models~1.3 and 1.4, this effect is added for both incidence and mortality, but in Model~1.3, the variance parameter is shared by both effects.

Model~3.1 is given by \autoref{Eq:2}, and Models~3.2 to 3.4 modify the spatial effect similarly to Models~1.2 to 1.4. Given that the number of scaling parameters used in shared spatio-temporal effects can vary, we consider different versions of the same model depending on the number of scaling parameters we have introduced.  To distinguish each case, the model labelling is differentiated using subscripts a ($l=1$), b ($l=3$ or $l=7$) or  c ($l=T$). In this section, we present results for a and b, while results for c are available in \autoref{AppendixC}. Firstly, for models with subindex a, we examine the case in which a single scaling parameter ($l = 1$) is included (representing the most restrictive model). Secondly, for models with subindex b, we select $l=3$ or $l=7$ scaling parameters depending on the cancer location. To select the number of scaling parameters for each cancer site, we analyze the results observed in the exploratory analysis and the results obtained with $l=T$ (models c, presented in \autoref{AppendixC}).
We finally take three different scaling parameters for pancreatic cancer, repeating each parameter over three periods. Additionally, we consider seven scaling parameters for leukaemia, with one assigned to each period, except for the 5th and 6th periods where the same scaling parameter is used, as well as the 7th and 8th periods where another identical scaling parameter is defined.  The complete description of the eight models can be found in \autoref{AppendixC}.

\autoref{tab2} presents the model selection criteria. It is evident that all multivariate spatio-temporal models incorporating shared spatio-temporal interaction terms outperform the multivariate models with specific interaction for each health outcome. Furthermore, in line with the findings from the exploratory data analysis, a distinct number of scaling parameters is chosen for each cancer site.  Similar results are obtained for pancreatic cancer  with one scaling parameter and three scaling parameters, then we select the simplest model  (Model~3.1a) to analyze pancreatic cancer. Regarding leukaemia, Model~3.2b ($l=7$) exhibits the best criteria values. Therefore, we select Model~3.2b to analyze leukaemia data.
Regarding computational cost, fitting Model~3.1a for pancreatic cancer takes less than one minute, while fitting Model~3.2b with 7 scaling parameters for leukaemia takes nearly 45 minutes.

\subsubsection{Pancreatic cancer}

\begin{figure}[t!]
	\begin{center}
			\scalebox{0.064}{\includegraphics[page=1]{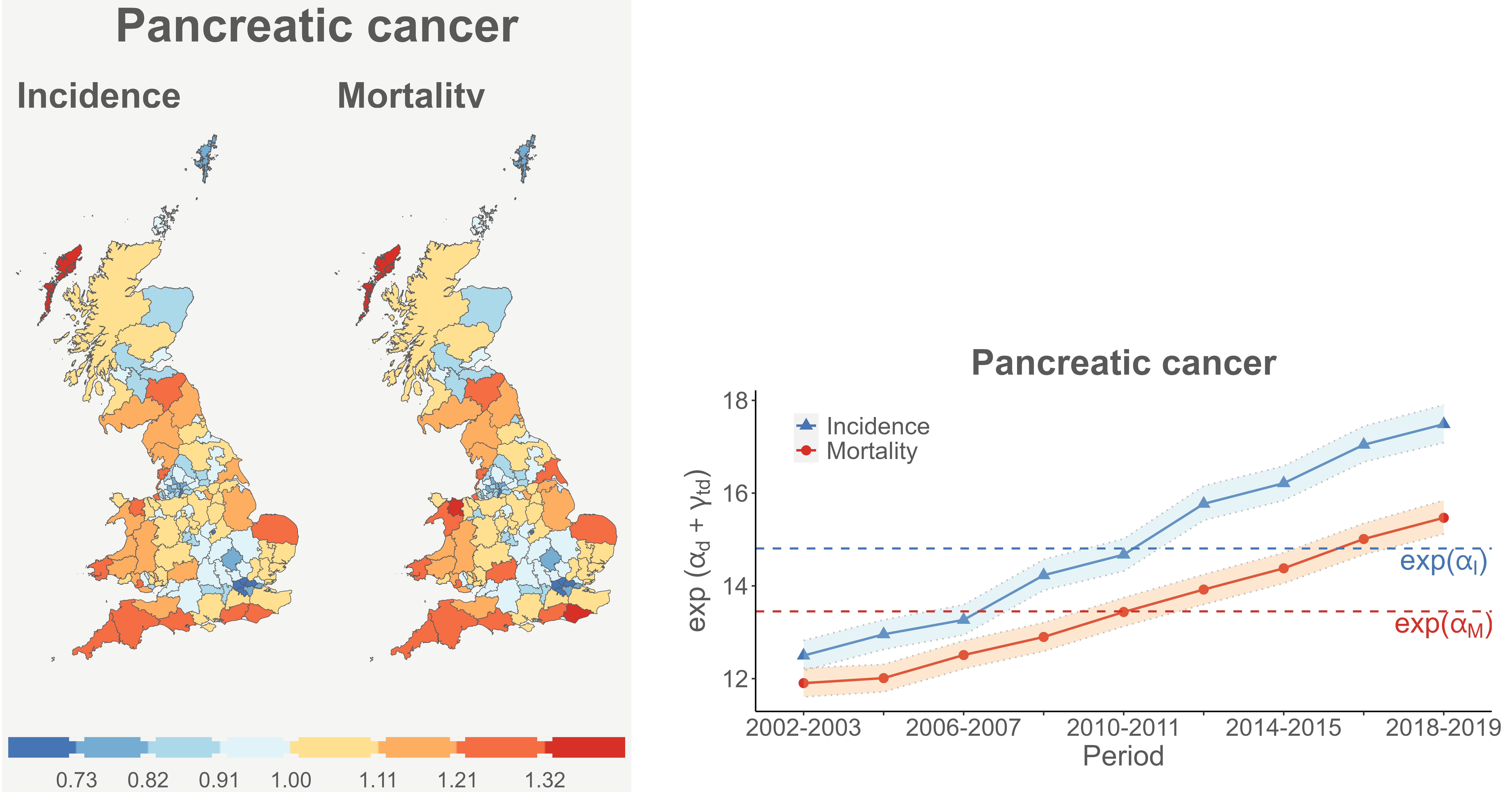}}
	\end{center}
	\caption{\label{fig4} Posterior median of the area-specific shared spatial effect, $\exp (\delta\kappa_{i})$ and $\exp \left(\frac{1}{\delta}\kappa_{i}\right)$ (left), and posterior median and 95\% credible interval of the health outcome-specific temporal component, $\exp (\alpha_d+\gamma_{td})$ (right) for pancreatic cancer.}
\end{figure}

The area-specific shared spatial effect for each cancer site, $\exp (\delta\kappa_i)$ or $\exp(\frac{1}{\delta}\kappa_i)$, captures the underlying common geographical pattern of incidence and mortality, respectively. This can be interpreted as a common spatial risk pattern 
that both health outcomes share, and it may reflect the effect of potential spatial risk factors such as certain demographic or socio-economic characteristics. The scaling parameter, $\delta$, determines the relationship between  the spatial pattern of cancer incidence and mortality, thereby increasing or decreasing the influence of the shared risk pattern for each health outcome. Posterior medians of the area-specific shared spatial effect for pancreatic cancer are displayed on the left side of \autoref{fig4}. We notice  that the cancer incidence and mortality shared spatial pattern is very similar ($\delta = 0.97$), although minor disparities can be observed in certain areas. Values greater than 1 indicate areas with rates exceeding the global rate of the health outcome ($\text{exp}\, \alpha_d$), while values less than 1 indicate areas with rates below the global rate. For both incidence and mortality, high values are observed in coastal areas, with some exceptions in eastern Scotland and areas in south Wales. The lowest values are observed in areas near London and Manchester.

The global temporal evolution of each health outcome in Great Britain is revealed by the health outcome-specific temporal component, $\exp (\alpha_d+\gamma_{td})$. This component helps determine whether specific events such as policy changes, shifts in government, or broader societal transformations have an impact on health outcomes over time.
\autoref{fig4} presents the posterior medians of $\exp (\alpha_d+\gamma_{td})$ along with their corresponding 95\% credible intervals, while the horizontal lines represent the global rates. Our results indicate a consistent upward trend in both incidence and mortality rates over the observed period, with mortality demonstrating a slower growth compared to incidence.

\begin{figure}[t!]
	\begin{center}
		\scalebox{0.048}{\includegraphics[page=1]{Fig/Fig6.jpeg}}
	\end{center}
	\caption{\label{fig5b} Posterior medians and 95\% credible interval of the spatio-temporal effect ($\exp (\alpha_I + \varrho \chi_{it})$ and $\exp \left(\alpha_M + \frac{1}{\varrho}\chi_{it}\right)$) for pancreatic incidence and mortality respectively, in Cardiff, Lothian (Edinburgh), NHS Liverpool CCG and NHS North East London CCG.}
\end{figure}

Area-specific temporal trends, that is, the posterior medians of $\exp(\alpha_I + \varrho\chi_{it})$ and $\exp(\alpha_M + \frac{1}{\varrho}\chi_{it})$, with 95\% credible intervals and the global rates ($\exp(\alpha_I)$ and $\exp(\alpha_M)$)  for pancreatic cancer in four selected  areas (Cardiff, Edinburgh, Liverpool and North East London) are shown in \autoref{fig5b}. In the case of pancreatic cancer, a  Type~I interaction with one scaling parameter was selected as the best model.
Consequently, as shown in \autoref{fig5b}, the temporal trends obtained for each area show a proportional relationship between incidence and mortality.

To conclude the analysis, we compute the evolution of the geographical distribution of the rates per \mbox{100 000} inhabitants. To save space, \autoref{figA2} in \autoref{AppendixD} shows the posterior medians of the rates per \mbox{100 000} inhabitants, $r_{itd}*10^5$, for pancreatic cancer. The maps reveal a noticeable rise in both incidence and mortality rates, with a similar pattern observed for both health outcomes. The regions with the highest estimated rates are concentrated in Wales and its surrounding areas, the coastal regions of southern England, the border areas between England and Scotland, and the coastal areas of western Scotland.

\subsubsection{Leukaemia}

\begin{figure}[t!]
	\begin{center}
		\scalebox{0.064}{\includegraphics[page=1]{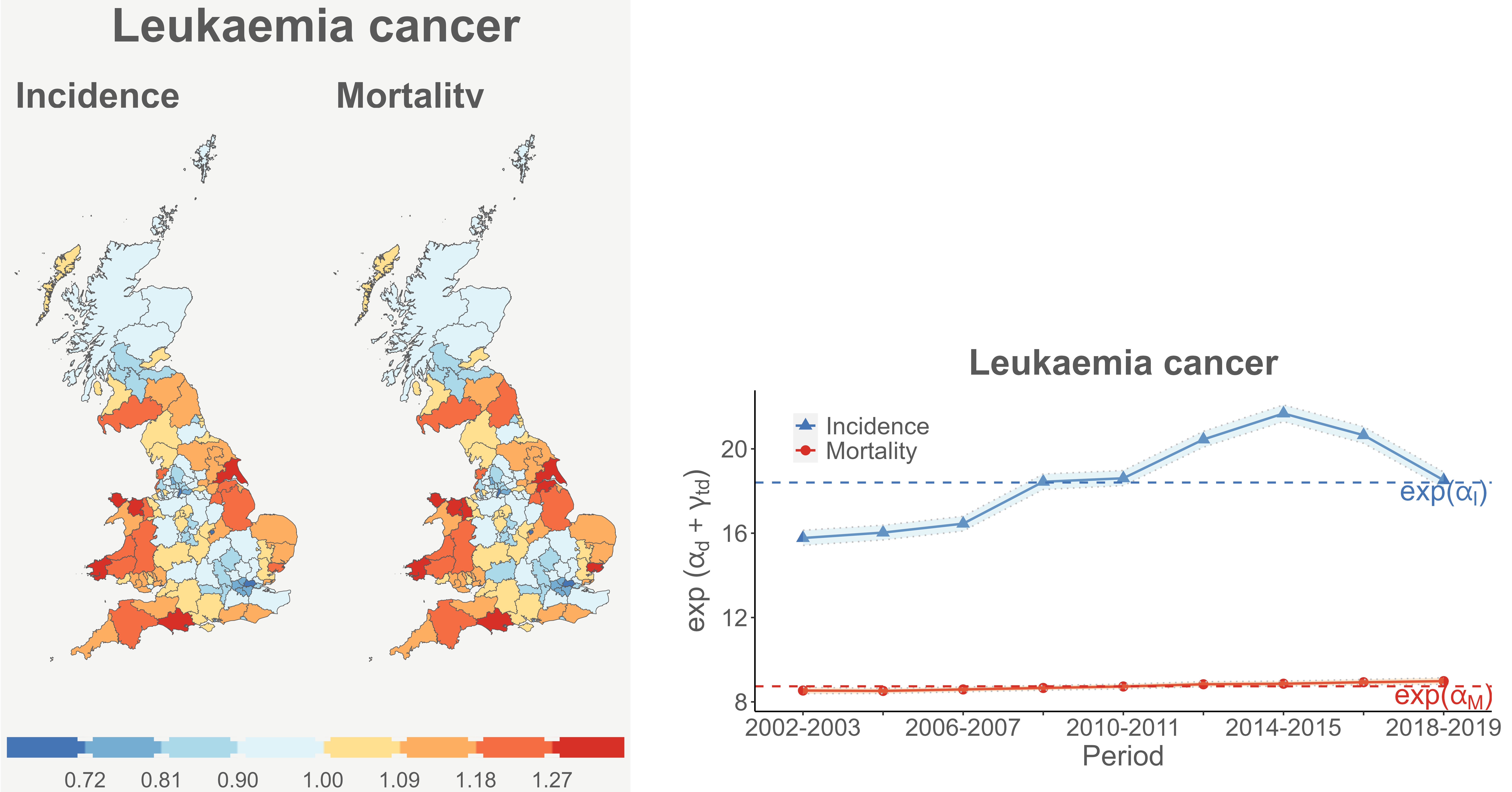}}
	\end{center}
	\caption{\label{fig5} Posterior median of the area-specific shared spatial effect, $\exp (\delta\kappa_{i})$ and $\exp \left(\frac{1}{\delta}\kappa_{i}\right)$ (left), and posterior median and 95\% credible interval of the health outcome-specific temporal component, $\exp \left(\alpha_d+\gamma_{td}\right)$ (right) for leukaemia cancer.}
\end{figure}

\autoref{fig5} on the left displays the posterior medians of the area-specific shared spatial effect for leukaemia cancer. As previously mentioned, this effect captures the common underlying geographical pattern of both incidence and mortality, allowing for the examination of potential shared spatial risk factors affecting both health outcomes. For leukaemia cancer, we obtain a value of $\delta = 0.99$, indicating a strong association between cancer incidence and mortality. \autoref{fig5} shows	high values in Wales, south and east coastal areas of England and the areas of Scotland neighbouring England. Low values are observed on most areas of Scotland and in areas near London and Manchester. Moreover, for leukaemia cancer the model selected has an spatially unstructured random effect for mortality which represents area-specific effects that can not be explained by the shared term and allows to identify potential risk factors affecting leukaemia cancer mortality but not incidence. \autoref{figA1} in \autoref{AppendixD} shows the posterior medians of the area-specific spatially unstructured random effect. Once again, values greater than 1 indicate areas with rates surpassing the global rate, whereas values less than 1 indicate areas with rates below the global rate. Low values are observed in Wales, areas located east from Manchester and areas close to London. In contrast, high values are observed in the coastal areas south of London and in the areas of England neighbouring Wales.

\autoref{fig5} on the right shows the global temporal evolution of each health outcome (posterior medians of $\exp (\alpha_d+\gamma_{td})$  and their 95\% credible intervals) and the global rates ($\text{exp} (\alpha_I)$ and $\text{exp} (\alpha_M)$).
The trend for leukaemia mortality remains relatively stable, showing a slight increase from 2002 to 2019. In contrast, the trend for leukaemia incidence shows slow growth for the first three years, followed by a faster increase that reaches the global incidence rate and remains stable for a year. Afterwards, an inverted U-shape emerges, with the maximum value recorded in 2014-2015.

\begin{figure}[t!]
	\begin{center}
		\scalebox{0.048}{\includegraphics[page=1]{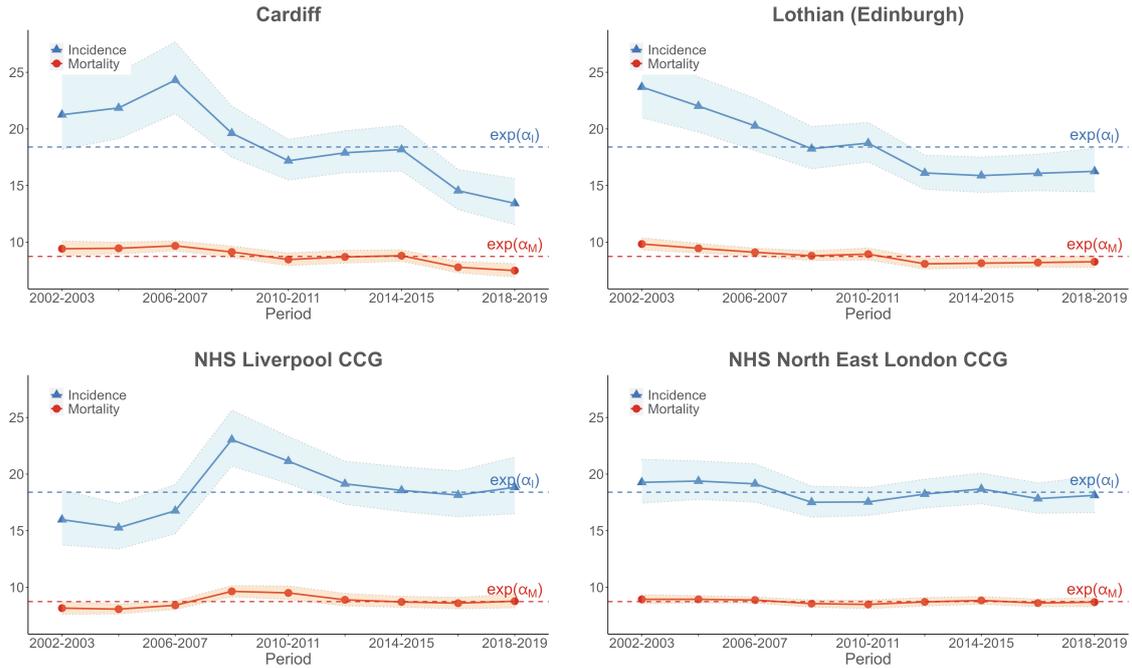}}
	\end{center}
	\caption{\label{fig7} Posterior medians and 95\% credible interval of the spatio-temporal effect ($\exp (\alpha_I + \varrho_t\chi_{it})$ and $\exp (\alpha_M + \frac{1}{\varrho_t}\chi_{it})$ ) for leukaemia incidence and mortality respectively, in Cardiff, Lothian (Edinburgh), NHS Liverpool CCG and NHS North East London CCG.}
\end{figure}

The shared interaction term, $\chi_{it}$, allows a different time evolution for each area that is shared for both incidence and mortality, but with different scaling parameter $\varrho_t$. For leukaemia, we select a model including a Type~IV interaction with 7 scaling parameters.
This indicates varying levels of association between disease incidence and mortality across the time period examined.
\autoref{fig7} displays the area-specific temporal trends of leukaemia cancer incidence and mortality for four regions (Cardiff, Edinburgh, Liverpool and North East London), presenting the posterior medians of $\exp(\alpha_I + \varrho_t\chi_{it})$   and $\exp(\alpha_M + \frac{1}{\varrho_t}\chi_{it})$ along with their 95\% credible intervals and the national rates.
The specific temporal trends in each area clearly differ, with some regions (such as Liverpool) showing an increase, while others (such as Cardiff, Edinburgh, and North East London) experience either a decrease or remain relatively stable.
Moreover, since different scaling parameters have been estimated (all with posterior medians greater than 1), incidence rates exhibit more pronounced changes compared to mortality rates.

\autoref{figA1*} in \autoref{AppendixD} shows the posterior medians of $\exp(\varrho_t\chi_{it})$ and $\exp(\frac{1}{\varrho_t}\chi_{it})$ for leukaemia incidence and mortality respectively. The evolution varies according to the geographical area analyzed, but in general, it is noticeable that in most areas of Scotland until 2010-2011 the values remain constant and then, a decrease is observed. In the case of Wales, the values remain constant until 2006-2007, then undergo a decrease for two periods, followed by an increase for two periods, and ultimately another decrease. In England, there are more disparities observed depending on the specific area being analyzed.

To conclude the analysis, \autoref{figA3} in \autoref{AppendixD} shows the evolution of the geographical distribution of incidence and mortality rates per \mbox{100 000} inhabitants, $r_{itd}*10^5$ (posterior medians). 
There is a distinct evolution for incidence and mortality. In the case of leukaemia incidence, an initial increase in rates is observed until the period 2014-2015, after which a decrease occurs. Conversely, mortality rates exhibit a relatively stable pattern across the studied periods. The areas with the most significant increase during the study period are located in Wales and England.

\section{Discussion}\label{Section4:Discussion}

In this work, we introduce a novel multivariate spatio-temporal model incorporating flexible shared spatio-temporal effects to jointly analyze incidence and mortality of rare cancers. Our approach is  built on the fundamental idea that when health outcomes have low rates, sharing spatio-temporal components can significantly enhance the accuracy of rate estimates. Our model is based on a combination of two established frameworks: the shared component models, which are suitable for cases where the relationship between health outcomes is known in advance, as it is the case of cancer incidence and mortality, and the explainable spatio-temporal interactions proposed by \cite{knorr2000}. The novelty of our proposal lies in the inclusion of a time-varying scaling parameter in the shared spatio-temporal effect. This model enables the modulation of the spatio-temporal spatio-temporal effects between cancer incidence and mortality, accommodating changes in their relationship over time.

Regarding model identifiability,  several considerations need to be addressed. Firstly, the scaling parameters of the flexible shared spatio-temporal effect are directly identifiable. This is because the mortality scaling parameter is defined as the inverse of the incidence scaling parameter at each specific time period, satisfying the constraint defined by \cite{held2005}. Secondly, since the model comprises intercepts, spatial, temporal, and spatio-temporal interactions, additional constraints are necessary. The specific constraints to be applied depend on the type of interaction incorporated in the selected model \citep{goicoa2018}. The models proposed in this work were fitted using integrated nested Laplace approximations in R-INLA. However, as these new models are not directly available in INLA,  we have developed our own implementation using the \texttt{rgeneric} model \citep{gomez2020}.

A simulation study has been conducted to analyze the behaviour of the novel multivariate spatio-temporal model with flexible shared spatio-temporal components. Three scenarios have been defined according to the different spatio-temporal terms outlined in this work. The results indicate that the new shared interactions with time-varying scaling parameter are adaptive enough to accommodate to all scenarios analyzed. These results enable us to recommend the use of the models proposed in this work to analyse rare or less frequent cancers. Furthermore, these new models have been employed to examine the spatio-temporal patterns of leukaemia and pancreatic cancer incidence and mortality rates in males within 142 administrative health care districts across Great Britain from 2002 to 2019. Model selection criteria indicate that these new models outperform conventional spatio-temporal models. Our real data analyses yield valuable insights for public health authorities, offering a comprehensive overview of the evolution of leukaemia and pancreatic cancer rates.

We would like to remark that studying incidence or mortality separately in rare cancers does not generally permit to estimate the specific temporal trend of a particular region. In other words, limited data hinders proper estimation of the spatio-temporal interaction. However, by considering incidence and mortality together we can analyze a region's specific behaviour over time.  Indeed, joint analysis of incidence and mortality rates, as proposed in this paper, offers valuable insights for health researchers.
In epidemiology, incidence tracks new disease cases within a population over time, aiding in identifying risk factors. Mortality, on the other hand, measures deaths from the disease within the same population and time frame, indicating disease severity and the effectiveness of ongoing health care interventions. When analyzed together, these metrics provide a comprehensive picture of disease dynamics, crucial for developing effective public health strategies.
This does not preclude that the models introduced here could be also used to analyzed other diseases that share common risk factors or even the same disease by sex.

Our proposal also permits to include covariates such as environmental or socio-demographic factors, that could explain changes in spatial and temporal patterns. Furthermore, this framework can be extended to other likelihoods by specifying a different likelihood in the INLA function and handle missing data, particularly when many spatial or temporal units do not report any data. When dealing with missing data in spatial or temporal units, it becomes challenging to estimate the spatial and temporal effects as we need to aggregate the data to estimate them. This is why the spatio-temporal  term provides the most disaggregation level of the data and, consequently, the most information to obtain predictions for missing data. Models with shared interactions with time-varying scaling parameters enhanced information sharing between cancer incidence and mortality, with the expectation of improving rate predictions.
This aspect is of significant interest to several European Population Based Cancer Registries, as can be seen in the work of \cite{retegui2023}, where they analyzed different multivariate models to predict missing data in a spatial context. It is interesting to extend that analysis to a spatio-temporal context, which will be addressed in future research.

	\section*{Acknowledgements}
	The work was supported by Project PID2020-113125RB-I00/MCIN/AEI/10.13039/501100011033, Project UNEDPAM/PI/PR24/05A and Ayudas Predoctorales Santander UPNA 2021-2022. 
	
	\section*{Competing Interests}
	All authors certify that they have no affiliations with or involvement in any organization or entity with any financial interest or non-financial interest in the subject matter or materials discussed in this manuscript.

	\section*{Code availability}
	The code and datasets is available online under \url{https://github.com/spatialstatisticsupna//Shared_interactions}.

	\bibliography{BIB}
	
	\begin{appendices}
	
		\section{Deriving the precision matrix}\label{AppendixA}
		
		We have implemented the flexible shared spatio-temporal models using \texttt{rgeneric} model. This model allows the user to define latent model components in R, and do the Bayesian inference using INLA. INLA performs approximate fully Bayesian inference of the class of latent Gaussian models \citep[LGMs,][]{rue2009}. Latent Gaussian models are statistical models that relate the response variable to an additive linear predictor while assuming a Gaussian Markov Random Field (GMRF) for the latent field of the model \citep{rue2005}. Therefore, a multivariate Gaussian prior with a sparse precision matrix is assumed for the latent field. When implementing a latent effect with \texttt{rgeneric} model, we encounter various necessary functions, including one that defines the precision matrix and another that determines the mean of the multivariate Gaussian distribution. For more information see \citet[][Section 11.3]{gomez2020}.
		When implementing the flexible shared component models, an additional drawback arises as INLA does not allow the repeated incorporation of a latent effect within the same model \citep{martins2013}, and in these particular models, we encounter the challenge that the interaction effect is common to both health outcomes. Therefore, to implement such models, we rely on the \texttt{copy} feature defined in R-INLA \citep{martins2013}. This feature enables us to incorporate the same latent effect twice in our model, by generating an almost identical copy of the latent field that is required multiple times in the model formulation. More precisely, to define the flexible shared component models, if we denote the latent effect of the spatio-temporal interactions by
		\begin{equation}
			\mathbf{z} = \boldsymbol{Z_3}\boldsymbol{\chi}, \nonumber
		\end{equation}
		we define an extended latent effect $\boldsymbol{x} = \left(\boldsymbol{z},\boldsymbol{z^*}\right)$ where $\boldsymbol{z^*}$ is the almost identical copy of $\boldsymbol{z}$. Moreover, it is also possible for the copied latent effect to have a scale parameter $\lambda$. Therefore, we define $\boldsymbol{z^*}$ as
		\begin{equation*}
			\boldsymbol{z^*} = \lambda \boldsymbol{z} + \boldsymbol{\epsilon}
		\end{equation*}
		where $\boldsymbol{\epsilon}$ is a tiny error that controls the degree of closeness between $\boldsymbol{z}$ and $\boldsymbol{z}^*$. In our context, we need to consider that the copied latent effect defined in the model is $\boldsymbol{Z_3}^{-1}\boldsymbol{\chi}$. As such, it is necessary that $\boldsymbol{Z_3}^{-1}\boldsymbol{\chi} = \lambda \boldsymbol{z}$, which implies that the value of the unknown scale parameter must be $\lambda = \left(\boldsymbol{Z_3}^{-1}\right)^2$. Therefore to implement the flexible shared component model we have defined the copied latent effect $\boldsymbol{z^*}$ as
		\begin{eqnarray}
			\mathbf{z^*} &=& \boldsymbol{Z_3}^{-1}\boldsymbol{\chi} + \boldsymbol{\epsilon} = \left(\boldsymbol{Z_3}^{-1}\right)^2\boldsymbol{z} + \boldsymbol{\epsilon}. \label{Eq:A5}
		\end{eqnarray}
		Additionally, the structure of $\boldsymbol{z}^*$ is inherited from $\boldsymbol{\epsilon}$ and hence we define a spatio-temporal structure for $\boldsymbol{\epsilon}$. Therefore, $\boldsymbol{\epsilon}$ follows a gaussian distribution with mean $\boldsymbol{0}$ and precision matrix $\tau_{\epsilon}\mathbf{Q}_{\chi}$. To achieve an almost identical copy of z, we set a high precision value, specifically, $\tau_{\epsilon} = \exp(15)$ \citep[see][]{martins2013}.
		
		To implement the extended latent effect $\boldsymbol{x}$ with the \texttt{rgeneric} model, we need to define the distribution of $\boldsymbol{x}$. Following the definition of the joint distribution of random variables, we obtain that the distribution of the new latent effect $\boldsymbol{x}$ is,
		\begin{equation}
			\pi\left(\boldsymbol{x}\right) = \pi\left(\boldsymbol{z}\right)\pi\left(\boldsymbol{z}^*|\boldsymbol{z}\right).\label{Eq:A6}
		\end{equation}
		Therefore, we are going to compute the distribution of $\boldsymbol{x}$ as the product of the distribution of the latent effect $\boldsymbol{z}$ and the distribution of the copied latent effect $\boldsymbol{z}^*$ conditional to $\boldsymbol{z}$.
		First, we compute the distribution of the latent effect $\boldsymbol{z}$. We have defined the distribution of the spatio-temporal interaction $\boldsymbol{\chi}$ as $p(\boldsymbol{\chi}) \propto \exp\left(\frac{-\tau_{\chi}}{2} \boldsymbol{\chi}^{'} \mathbf{Q}_{\chi} \boldsymbol{\chi} \right)$.
		That is, $\boldsymbol{\chi}$ follows a gaussian multivariate distribution with mean $\boldsymbol{0}$ and precision matrix $\tau_{\chi}\mathbf{Q}_{\chi}$.
		We have defined $\boldsymbol{z} = \boldsymbol{Z_3}\boldsymbol{\chi}$, i.e. $\boldsymbol{z}$ is the product of a matrix of real values $\boldsymbol{\varrho}$ and a random variable that follows a gaussian multivariate distribution. Following \citet[][Section~3.2.]{tong2012}, we obtain that $\boldsymbol{z}$ follows a multivariate gaussian distribution. Specifically,
		\begin{equation}
			\boldsymbol{z} \sim N\left(\boldsymbol{0},\boldsymbol{Z_3}\left(\frac{1}{\tau_{\chi}}\mathbf{Q}_{\chi}^-\right)\boldsymbol{Z_3}\right). \label{Eq:8}
		\end{equation}
		Thus, the precision matrix is ${\Sigma^- = \boldsymbol{Z_3}^{-1}\left(\tau_{\chi}\mathbf{Q}_{\chi}\right)\boldsymbol{Z_3}^{-1}}$.
		
		Now, we compute the distribution of $\boldsymbol{z}^*$ conditional to $\boldsymbol{z}$, i.e. $\pi\left(\boldsymbol{z}^*|\boldsymbol{z}\right)$. Remember that we have defined $\boldsymbol{z}^*$ as a linear combination of $\boldsymbol{z}$ and a tiny error $\boldsymbol{\epsilon}$ (see \autoref{Eq:A5}) and $\boldsymbol{\epsilon}$ follows a gaussian distribution with mean $\boldsymbol{0}$ and a high precision. Specifically,
		\begin{equation*}
			\boldsymbol{\epsilon} \sim N\left(\boldsymbol{0},\frac{1}{\tau_{\epsilon}}\boldsymbol{Q}_\chi^-\right).
		\end{equation*}
		
		Recall that a linear function of a multivariate normal is itself a multivariate normal distribution \citep[see][Section~3]{tong2012}. Since the latent effect $\boldsymbol{z}$ is known, $\boldsymbol{z}^*$ conditional on $\boldsymbol{z}$ is just a multivariate normal random variable. If we compute the mean and the variance of the multivariate normal distribution of $\boldsymbol{z}^*$ conditional to $\boldsymbol{z}$, we obtain:
		\begin{eqnarray*}
			\mu &=& E\left(\boldsymbol{z}^*\right) = E\left(\left(\boldsymbol{Z_3}^{-1}\right)^2\boldsymbol{z}+\boldsymbol{\epsilon}\right) = E\left(\left(\boldsymbol{Z_3}^{-1}\right)^2\boldsymbol{z}\right) + E\left(\boldsymbol{\epsilon}\right) = \left(\boldsymbol{Z_3}^{-1}\right)^2\boldsymbol{z} + \boldsymbol{0} \\
			&=& \left(\boldsymbol{Z_3}^{-1}\right)^2\boldsymbol{z}\\
			\Sigma_{z^*} &=& Var\left(\boldsymbol{z}^*\right) = Var\left(\left(\boldsymbol{Z_3}^{-1}\right)^2\boldsymbol{z}+\boldsymbol{\epsilon}\right) = Var\left(\left(\boldsymbol{Z_3}^{-1}\right)^2\boldsymbol{z}\right) + Var\left(\boldsymbol{\epsilon}\right) =  \boldsymbol{0} + \frac{1}{\tau_{\epsilon}}\boldsymbol{Q}_\chi^- \\
			&=& \frac{1}{\tau_{\epsilon}}\boldsymbol{Q}_\chi^-
		\end{eqnarray*}
		Then, the distribution of $\boldsymbol{z}^*$ conditional to $\boldsymbol{z}$ is as follows,
		\begin{eqnarray}
			\boldsymbol{z}^*|\boldsymbol{z} & \sim & N\left(\mu,\Sigma_{z^*}\right)= N\left(\left(\boldsymbol{Z_3}^{-1}\right)^2\boldsymbol{z},\frac{1}{\tau_{\epsilon}}\boldsymbol{Q}_\chi^-\right).
			\label{Eq:12}
		\end{eqnarray}
	
		Consequently, we can replace the probability density functions defined by the distributions obtained in \autoref{Eq:8} and \autoref{Eq:12} in \autoref{Eq:A6} to be able to compute the distribution of $\boldsymbol{x}$:
		{\small \begin{eqnarray}\label{equation:distx}
			\pi\left(\boldsymbol{x}\right) &=& \pi\left(\boldsymbol{z}\right)\pi\left(\boldsymbol{z}^*|\boldsymbol{z}\right) \nonumber\\
			& \propto& \exp\left(-\frac{1}{2}\boldsymbol{z}^{'}\left(\boldsymbol{Z_3}^{-1}\left(\tau_{\chi}\mathbf{Q}_{\chi}\right)\boldsymbol{Z_3}^{-1}\right)\boldsymbol{z}\right) \exp\left(-\frac{\tau_{\epsilon}}{2}(\boldsymbol{z}^*-\left(\boldsymbol{Z_3}^{-1}\right)^2\boldsymbol{z})^{'}\boldsymbol{Q}_{\chi}(\boldsymbol{z}^*-\left(\boldsymbol{Z_3}^{-1}\right)^2\boldsymbol{z})\right) \nonumber\\
			&\propto& \exp\left(-\frac{1}{2}\left[\tau_{\chi}\boldsymbol{z}^{'}\left(\boldsymbol{Z_3}^{-1}\mathbf{Q}_{\chi}\boldsymbol{Z_3}^{-1}\right)\boldsymbol{z} + \tau_{\epsilon} (\boldsymbol{z}^*-\left(\boldsymbol{Z_3}^{-1}\right)^2\boldsymbol{z})^{'}\boldsymbol{Q}_{\chi}(\boldsymbol{z}^*-\left(\boldsymbol{Z_3}^{-1}\right)^2\boldsymbol{z}) \right]\right)
		\end{eqnarray}}
	
		We need to develop parts $\boldsymbol{z}^{'}\left(\boldsymbol{Z_3}^{-1}\mathbf{Q}_{\chi}\boldsymbol{Z_3}^{-1}\right)\boldsymbol{z}$ and $(\boldsymbol{z}^*-\left(\boldsymbol{Z_3}^{-1}\right)^2\boldsymbol{z})^{'}\boldsymbol{Q}_{\chi}(\boldsymbol{z}^*-\left(\boldsymbol{Z_3}^{-1}\right)^2\boldsymbol{z})$ of the distribution of $\boldsymbol{x}$. We start by developing $\boldsymbol{z}^{'}\left(\boldsymbol{Z_3}^{-1}\mathbf{Q}_{\chi}\boldsymbol{Z_3}^{-1}\right)\boldsymbol{z}$. We define $\boldsymbol{Q^*} = \boldsymbol{Z_3}^{-1}\mathbf{Q}_{\chi}\boldsymbol{Z_3}^{-1}$ as
		
		{\small \begin{equation}
			\boldsymbol{Q^*} = \boldsymbol{Z_3}^{-1}\mathbf{Q}_{\chi}\boldsymbol{Z_3}^{-1} = \left( \begin{array}{cccc}
				{Z_3}^{-1}_{11}Q^\chi_{11}{Z_3}^{-1}_{11} & {Z_3}^{-1}_{11}Q^\chi_{12}{Z_3}^{-1}_{22} & \dots & {Z_3}^{-1}_{11}Q^\chi_{1,TA}{Z_3}^{-1}_{TA,TA}\\
				{Z_3}^{-1}_{22}Q^\chi_{21}{Z_3}^{-1}_{11} & {Z_3}^{-1}_{22}Q^\chi_{22}{Z_3}^{-1}_{22} & \dots & {Z_3}^{-1}_{22}Q^\chi_{2,TA}{Z_3}^{-1}_{TA,TA}\\
				\vdots & \vdots & & \vdots\\
				{Z_3}^{-1}_{TA,TA}Q^\chi_{TA,1}{Z_3}^{-1}_{11} & {Z_3}^{-1}_{TA,TA}Q^\chi_{TA,2}{Z_3}^{-1}_{22} & \dots & {Z_3}^{-1}_{TA,TA}Q^\chi_{TA,TA}{Z_3}^{-1}_{TA,TA}\\
			\end{array}
			\right)\nonumber
		\end{equation}}
		
		Then, defining as $Q^*_{ij}$ the elements and $Q^*_{.j}$ the $j$th column of $\boldsymbol{Q^*}$ we obtain for $\boldsymbol{z}^{'}\left(\boldsymbol{Z_3}^{-1}\mathbf{Q}_{\chi}\boldsymbol{Z_3}^{-1}\right)\boldsymbol{z}$:
		
		\begin{eqnarray}\label{equation:distxpart1}
			\boldsymbol{z}^{'}\left(\boldsymbol{Z_3}^{-1}\mathbf{Q}_{\chi}\boldsymbol{Z_3}^{-1}\right)\boldsymbol{z} &=& \boldsymbol{z}^{'}\boldsymbol{Q^*}\boldsymbol{z} =
			\left(\boldsymbol{z}^{'}Q^*_{.1} \quad \boldsymbol{z}^{'}Q^*_{.2} \quad \dots \quad \boldsymbol{z}^{'}Q^*_{.n}\right)\boldsymbol{z} \nonumber \\
			&=& \left(\sum_{i=1}^{TA}z_iQ^*_{i1} \quad \sum_{i=1}^{TA}z_iQ^*_{i2} \quad \dots \quad \sum_{i=1}^{TA}z_iQ^*_{in} \right) \boldsymbol{z}  \nonumber \\
			&=& \sum_{j=1}^{TA} z_j \sum_{i=1}^{TA}z_iQ^*_{ij} = \sum_{j=1}^{TA} \sum_{i=1}^{TA}z_j z_iQ^*_{ij} \nonumber \\
			&=& \sum_{j=1}^{TA} \sum_{i=1}^{TA} z_jz_i\left({Z_3}_{ii}^{-1}Q^{\chi}_{ij}{Z_3}_{jj}^{-1}\right).
		\end{eqnarray}
		
		To develop ${(\boldsymbol{z}^*-\left(\boldsymbol{Z_3}^{-1}\right)^2\boldsymbol{z})^{'}\boldsymbol{Q}_{\chi}(\boldsymbol{z}^*-\left(\boldsymbol{Z_3}^{-1}\right)^2\boldsymbol{z})}$, first we define $\boldsymbol{z}^*-\left(\boldsymbol{Z_3}^{-1}\right)^2\boldsymbol{z}$ as
		\begin{equation}
			\boldsymbol{z}^*-\left(\boldsymbol{Z_3}^{-1}\right)^2\boldsymbol{z} = \left(\begin{array}{c}
				z^*_1 - \left({Z_3}^{-1}_{11}\right)^2z_1\\
				z^*_2 - \left({Z_3}^{-1}_{22}\right)^2z_2\\
				\vdots\\
				z^*_{TA} - \left({Z_3}^{-1}_{TA,TA}\right)^2z_{TA}
			\end{array}\right) \nonumber
		\end{equation}
		
		Following the result reached for ${\boldsymbol{z}^{'}\left(\boldsymbol{Z_3}^{-1}\mathbf{Q}_{\chi}\boldsymbol{Z_3}^{-1}\right)\boldsymbol{z}}$ in \autoref{equation:distxpart1}, for \linebreak ${(\boldsymbol{z}^*-\left(\boldsymbol{Z_3}^{-1}\right)^2\boldsymbol{z})^{'}\boldsymbol{Q}_{\chi}(\boldsymbol{z}^*-\left(\boldsymbol{Z_3}^{-1}\right)^2\boldsymbol{z})}$ we obtain: 	
		{\small \begin{eqnarray}\label{equation:distxpart2}
				(\boldsymbol{z}^*-\left(\boldsymbol{Z_3}^{-1}\right)^2\boldsymbol{z})^{'}\boldsymbol{Q}_{\chi}(\boldsymbol{z}^*-\left(\boldsymbol{Z_3}^{-1}\right)^2\boldsymbol{z}) &=& \sum_{j=1}^{TA} \sum_{i=1}^{TA} \left(z^*_j - \left({Z_3}^{-1}_{jj}\right)^2z_j\right)\left(z^*_i - \left({Z_3}^{-1}_{ii}\right)^2z_i\right)Q^\chi_{ij} \nonumber \\
				&=& \sum_{j=1}^{TA} \sum_{i=1}^{TA} \left(z^*_jz^*_i - z^*_j\left({Z_3}^{-1}_{ii}\right)^2z_i - \left({Z_3}^{-1}_{jj}\right)^2z_jz^*_i \right. \nonumber\\
				&& \left. + \left({Z_3}^{-1}_{jj}\right)^2z_j\left({Z_3}^{-1}_{ii}\right)^2z_i\right) Q^\chi_{ij}  \nonumber\\
				&=& \sum_{j=1}^{TA} \sum_{i=1}^{TA} \left({Z_3}^{-1}_{jj}\right)^2z_j\left({Z_3}^{-1}_{ii}\right)^2z_iQ^\chi_{ij} - \sum_{j=1}^{TA} \sum_{i=1}^{TA} z^*_j\left({Z_3}^{-1}_{ii}\right)^2 Q^\chi_{ij} \nonumber\\
				& &  - \sum_{j=1}^{TA} \sum_{i=1}^{TA} \left({Z_3}^{-1}_{jj}\right)^2z_jz^*_i Q^\chi_{ij} + \sum_{j=1}^{TA} \sum_{i=1}^{TA} z^*_jz^*_i Q^\chi_{ij}.
		\end{eqnarray}}
		
		Then, if we replace \autoref{equation:distxpart1} and \autoref{equation:distxpart2} in  \autoref{equation:distx}, we obtain for the distribution of $\boldsymbol{x}$:
		{\small \begin{eqnarray}\label{equation:distx2}
			\pi\left(\boldsymbol{x}\right) &\propto& \exp\left(-\frac{1}{2}\left[\tau_{\chi}\boldsymbol{z}^{'}\left(\boldsymbol{Z_3}^{-1}\mathbf{Q}_{\chi}\boldsymbol{Z_3}^{-1}\right)\boldsymbol{z} + \tau_{\epsilon} (\boldsymbol{z}^*-\left(\boldsymbol{Z_3}^{-1}\right)^2\boldsymbol{z})^{'}\boldsymbol{Q}_{\chi}(\boldsymbol{z}^*-\left(\boldsymbol{Z_3}^{-1}\right)^2\boldsymbol{z}) \right]\right)  \nonumber \\
			&\propto& \exp\left(-\frac{1}{2}\left[\tau_{\chi}\sum_{j=1}^{TA} \sum_{i=1}^{TA} z_jz_i\left({Z_3}_{ii}^{-1}Q^\chi_{ij}{Z_3}_{jj}^{-1}\right)  + \tau_{\epsilon}\left[\sum_{j=1}^{TA} \sum_{i=1}^{TA} \left({Z_3}^{-1}_{jj}\right)^2z_j\left({Z_3}^{-1}_{ii}\right)^2z_iQ^\chi_{ij}  \right.\right.\right. \nonumber \\
			& & \left.\left.\left. - \sum_{j=1}^{TA} \sum_{i=1}^{TA} z^*_j\left({Z_3}^{-1}_{ii}\right)^2 Q^\chi_{ij} - \sum_{j=1}^{TA} \sum_{i=1}^{TA} \left({Z_3}^{-1}_{jj}\right)^2z_jz^*_i Q^\chi_{ij} + \sum_{j=1}^{TA} \sum_{i=1}^{TA} z^*_jz^*_i Q^\chi_{ij}\right]\right]\right)  \nonumber \\
			&\propto& \exp\left(-\frac{1}{2}\left[\sum_{j=1}^{TA} \sum_{i=1}^{TA}  \left(\tau_{\chi}\left({Z_3}_{ii}^{-1}Q^\chi_{ij}{Z_3}_{jj}^{-1}\right)  + \tau_{\epsilon} \left({Z_3}^{-1}_{jj}\right)^2\left({Z_3}^{-1}_{ii}\right)^2Q^\chi_{ij}\right) z_jz_i \right.\right. \nonumber \\
			& & \left.\left. - \tau_{\epsilon}\sum_{j=1}^{TA} \sum_{i=1}^{TA} z^*_j\left({Z_3}^{-1}_{ii}\right)^2 Q^\chi_{ij} - \tau_{\epsilon}\sum_{j=1}^{TA} \sum_{i=1}^{TA} \left({Z_3}^{-1}_{jj}\right)^2z_jz^*_i Q^\chi_{ij} + \tau_{\epsilon}\sum_{j=1}^{TA} \sum_{i=1}^{TA} z^*_jz^*_i Q^\chi_{ij}\right]\right)  \nonumber \\
			&\propto& \exp\left(-\frac{1}{2}\boldsymbol{x}^T\boldsymbol{Q_x}\boldsymbol{x}\right)\nonumber
		\end{eqnarray}}
		where
		\begin{equation}
			\boldsymbol{Q_x} = \left(\begin{array}{cc}
				\tau_{\chi}\boldsymbol{Z_3}^{-1}\boldsymbol{Q}_\chi\boldsymbol{Z_3}^{-1}+\tau_{\epsilon}\left(\boldsymbol{Z_3}^{-1}\right)^2\boldsymbol{Q}_\chi\left(\boldsymbol{Z_3}^{-1}\right)^2& -\tau_{\epsilon}\left(\boldsymbol{Z_3}^{-1}\right)^2\boldsymbol{Q}_\chi\\
				-\tau_{\epsilon}\boldsymbol{Q}_\chi\left(\boldsymbol{Z_3}^{-1}\right)^2 & \tau_{\epsilon}\boldsymbol{Q}_\chi
			\end{array}\right). \nonumber
		\end{equation}
		Therefore, the distribution of $\boldsymbol{x}$ is
		\begin{equation}
			\boldsymbol{x} \sim N\left(\boldsymbol{0},\boldsymbol{Q_x}^-\right). \nonumber
		\end{equation}
		
		\newpage
		\section{Simulation Study}\label{AppendixB}
		
		This section shows additional tables that were discussed but not included in the main paper due to space limitations.
		
		\begin{table}[h!]
			\centering
			\caption{Percentiles of WAIC difference between the true model and the other models.  Symbol -  indicates that no differences are provided since it represents the true model.} 
			\resizebox{0.82\textwidth}{!}{
			\begin{tabular}{lrrrcrrrcrrr}
				\toprule
				&\multicolumn{3}{c}{Scenario 1} & & \multicolumn{3}{c}{Scenario 2} & & \multicolumn{3}{c}{Scenario 3}\\
				\midrule
				& \multicolumn{3}{c}{WAIC} && \multicolumn{3}{c}{WAIC}&& \multicolumn{3}{c}{WAIC}\\
				\midrule
				& \%2.5 & \%50 & \%97.5  && \%2.5 & \%50 & \%97.5 && \%2.5 & \%50 & \%97.5 \\
				&\multicolumn{11}{l}{Type I}\\		
				Model 1 & -  &  - & - &&  90.03 & 154.50 & 233.34  && 87.99 & 188.46 & 262.28  \\ 
				Model 2 &  7.61 & 33.97 & 62.29 && - &  - & -  && 42.96 & 95.98 & 134.92 \\ 
				Model 3 &  5.38 & 40.16 & 78.79& &  -5.82 & 5.13 & 12.33  && -  & -  &  -  \\ 
				\multicolumn{12}{l}{ }\\	
				&\multicolumn{11}{l}{Type II}\\	
				Model 1 &-   &  - &  -   &&  96.59 & 166.91 & 215.28  &&  138.24 & 206.77 & 284.90 \\ 
				Model 2 &  23.24 & 61.56 & 112.95  &&  - & - &-&& 63.64 & 107.01 & 167.36 \\ 
				Model 3 &28.95 & 73.95 & 127.44  &&  -7.04 & 2.80 & 7.46  && -  &  - &   -  \\ 
				\multicolumn{12}{l}{ }\\	
				&\multicolumn{11}{l}{Type III}\\	
				Model 1 & -  & -  & -   && 22.88 & 64.2 & 116.46 &&  38.84 & 85.56 & 136.61 \\ 
				Model 2 &  -0.37 & 11.44 & 35.73  &&  -  & -  & -  && 11.29 & 37.89 & 68.82\\
				Model 3 & 0.20 & 13.09 & 34.17 && -8.25 & 4.97 & 9.73 && - & -  & -   \\ 
				\multicolumn{12}{l}{ }\\	
				&\multicolumn{11}{l}{Type IV}\\	
				Model 1 & -  & -  &   - && 33.54 & 78.99 & 112.15  &&  50.52 & 95.15 & 151.88\\ 
				Model 2 & 7.11 & 26.34 & 52.4  &&  -  &  - & -  && 20.18 & 48.20 & 83.50 \\ 
				Model 3 & 14.87 & 36.26 & 60.77  && -8.14 & 2.03 & 8.91  &&- &- &-\\ 
				\bottomrule
			\end{tabular}
		}
		\end{table}
		
		\begin{sidewaystable}[h!]
			\centering
			\caption{MRRMSE values, together with the percentage change relative to the true model in each simulation study, the 95\% credible interval length (CIL) and the coverage percentage (Cov). Symbol -  indicates that no differences are provided since it represents the true model.} 
			\resizebox{0.82\textwidth}{!}{
			\begin{tabular}{lrrrrcrrrrcrrrr}
				\toprule
				&\multicolumn{4}{c}{Scenario 1} & & \multicolumn{4}{c}{Scenario 2} & & \multicolumn{4}{c}{Scenario 3}\\
				\midrule
				& MRRMSE & $\Delta^{MRRMSE}$& CIL & Cov  && MRRMSE & $\Delta^{MRRMSE}$& CIL & Cov &&  MRRMSE & $\Delta^{MRRMSE}$& CIL & Cov \\
				\midrule
				&\multicolumn{14}{l}{Type I}\\		
				Model 1 &  22.04 & -  & 2.77  & 94.94  &&  28.05 & 5.53 & 4.65 & 95.36 &&  28.33 & 6.22 & 4.80 & 95.35 \\ 
				Model 2 &  22.19 & 0.68 & 2.37 & 88.77  &&   26.58 & -  & 4.18 & 95.03 &&  27.16 & 1.84 & 4.24 & 93.49  \\ 
				Model 3 &  22.21 & 0.77 &  2.36 & 89.28 && 26.59 & 0.04 & 4.19 & 95.07 &&  26.67 & -  & 4.21 &  95.07 \\ 
				\multicolumn{15}{l}{ }\\	
				&\multicolumn{14}{l}{Type II}\\	
				Model 1 &   22.47 & -  & 2.84 & 94.61  &&  27.54 & 6.83 & 4.40 & 95.05 &&   28.37 & 6.61 & 4.54 & 93.90 \\ 
				Model 2 &  22.90 & 1.91 & 2.48 & 87.40 &&   25.78 & -  & 3.91 & 94.83   &&  27.12 & 1.92 & 3.95 & 91.53 \\ 
				Model 3 & 22.93 & 2.05 & 2.48 & 88.05  && 25.80 & 0.08 & 3.91 & 94.85  &&   26.61 & -  & 3.95 & 92.85 \\ 
				\multicolumn{15}{l}{ }\\	
				&\multicolumn{14}{l}{Type III}\\	
				Model 1 & 20.32 & -  & 2.32 & 94.92   && 24.40 & 3.52 & 3.44 & 94.68 &&  24.61 & 4.06 & 3.56 & 94.80 \\ 
				Model 2 & 20.47 & 0.74 & 2.09 & 91.40 &&  23.57 & -  & 3.24 & 94.83   && 24.01 & 1.52 & 3.31 & 93.63 \\
				Model 3 &  20.49 & 0.84 & 2.09 &  91.64 &&  23.61 & 0.17 & 3.24 &  94.75 && 23.65 & -  & 3.27 & 94.88  \\ 
				\multicolumn{15}{l}{ }\\	
				&\multicolumn{14}{l}{Type IV}\\	
				Model 1 &  20.66 & -  & 2.45 & 95.13   &&  24.45 & 4.58 & 3.48 & 95.03 && 25.05 & 4.59 & 3.58 & 94.48\\ 
				Model 2 &  21.01 & 1.69 & 2.24 & 90.48 && 23.38 & -  & 3.17 &  94.61  && 24.34 & 1.63 & 3.23 & 92.87 \\ 
				Model 3 &  21.10 & 2.13 & 2.22 & 90.46 &&  23.40 & 0.09 & 3.17 & 94.62 &&  23.95  &- & 3.23 & 94.04\\ 
				\bottomrule
			\end{tabular}
		}
		\end{sidewaystable}
		
		\clearpage	
		\section{Description of the multivariate spatio-temporal models}\label{AppendixC}
		
		We implemented eight different multivariate spatio-temporal models to carry out a joint pancreatic cancer and leukaemia study of incidence and mortality for males during the period 2002-2019 in Great Britain.
		
		To model the log rates, $\log r_{itd}$, we first consider four different spatio-temporal multivariate models all of them with a fixed effect for each health outcome, a shared spatial component, a time effect specific for each health outcome and independent interactions among health outcomes. Disparities among the models are seen in the spatial effect. Model~1.1 is the model defined in Equation~1 of the main paper and Models~1.2 to 1.4 modify the spatial effect by adding different spatially unstructured random effects. Precisely, for Model~1.2 we add a spatially unstructured random effect for mortality, for Model~1.3 and Model~1.4 the spatially unstructured random effect has been added for both incidence and mortality rates, but in Model~1.3 the variance parameter is shared by both effects. Therefore, we assume that the log rates, $\log r_{itd}$, have the decomposition
		\begin{eqnarray}
			&\textrm{Model 1.1:}&\log r_{itI} =\alpha_I + \delta \kappa_{i} + \gamma_{tI} + \chi_{itI}, \nonumber \\
			&&\log r_{itM} =\alpha_M + \frac{1}{\delta} \kappa_{i} + \gamma_{tM} + \chi_{itM}, \nonumber \\
			&&\nonumber \\
			&\textrm{Model 1.2:}&\log r_{itI} =\alpha_I + \delta \kappa_{i} + \gamma_{tI} + \chi_{itI}, \nonumber \\
			&&\log r_{itM} =\alpha_M + \frac{1}{\delta} \kappa_{i} + u_i + \gamma_{tM} + \chi_{itM}, \nonumber \\
			&&\nonumber \\
			&\textrm{Model 1.3:}&\log r_{itI} =\alpha_I + \delta \kappa_{i} + w_{iI} + \gamma_{tI} + \chi_{itI}, \nonumber \\
			&&\log r_{itM} =\alpha_M + \frac{1}{\delta} \kappa_{i} + w_{iM}+ \gamma_{tM} + \chi_{itM}, \nonumber \\
			&&\nonumber \\
			&\textrm{Model 1.4:}&\log r_{itI} =\alpha_I + \delta \kappa_{i} + v_i + \gamma_{tI} + \chi_{itI}, \nonumber \\
			&&\log r_{itM} =\alpha_M + \frac{1}{\delta} \kappa_{i} + u_i + \gamma_{tM} + \chi_{itM}, \nonumber
		\end{eqnarray}
		where $\alpha_d$ is a health outcome-specific intercept, $\delta$ is a scaling parameter, $\kappa_{i}$ represents the shared spatial component, $u_{i}$ represents the mortality specific spatially unstructured random effect, $w_{id}$ is a health outcome-specific spatially unstructured random effect, $v_i$ represents the incidence specific spatially unstructured random effect, $\gamma_{td}$ represents the time effect specific for each health outcome $d$ and $\chi_{itd}$ are the spatio-temporal interactions specific for each health outcome $d$. The priors for $\alpha_d$, $\delta$, $\kappa_{i}$, $\gamma_{td}$ and $\chi_{itd}$ can be found in the main paper. We assign the following priors to the spatially unstructured random effects
		
		\begin{eqnarray*}
			&&\boldsymbol{u}\sim N\left(\boldsymbol{0}, \tau_u\boldsymbol{I_A}\right),\\
			&&\boldsymbol{w}\sim N\left(\boldsymbol{0},\tau_{w}\left(\boldsymbol{I_2}\otimes \boldsymbol{I_A}\right)\right),\\
			&&\boldsymbol{v}\sim N\left(\boldsymbol{0}, \tau_v\boldsymbol{I_A}\right). 			
		\end{eqnarray*}

		Moreover, we propose a set of models with shared interactions among the health outcomes analyzed with the idea of improving estimates of health outcomes with low rates, since the amount of information shared is greater than in models with independent spatio-temporal interactions.
		To do so, we maintain the shared component model for area and the time effect for each health outcome $d$ as in the previous section, however, we define a shared component model for the interactions and as in the previous models we define different spatially unstructured random effects for each model. Precisely, Model~3.1 is the model defined in Equation~3 of the main paper and models 3.2 to 3.4 modify the spatial effect just like models 1.2 to 1.4. In this case, we assume that the log rates, $\log r_{itd}$, have the decomposition
		\begin{eqnarray}
			&\textrm{Model 3.1:}&\log r_{itI} = \alpha_I + \delta \kappa_{i} + \gamma_{tI} + \varrho_t \chi_{it}, \nonumber \\
			&&\log r_{itM} =\alpha_M + \frac{1}{\delta} \kappa_{i} + \gamma_{tM} + \frac{1}{\varrho_t} \chi_{it}, \nonumber \\
			&&\nonumber \\
			&\textrm{Model 3.2:}&\log r_{itI} = \alpha_I + \delta \kappa_{i} + \gamma_{tI} + \varrho_t \chi_{it}, \nonumber \\
			&&\log r_{itM} =\alpha_M + \frac{1}{\delta} \kappa_{i} + u_i + \gamma_{tM} + \frac{1}{\varrho_t} \chi_{it}, \nonumber \\
			&&\nonumber \\
			&\textrm{Model 3.3:}&\log r_{itI} = \alpha_I + \delta \kappa_{i} + w_{iI} + \gamma_{tI} + \varrho_t \chi_{it}, \nonumber \\
			&&\log r_{itM} =\alpha_M + \frac{1}{\delta} \kappa_{i} + w_{iM} + \gamma_{tM} + \frac{1}{\varrho_t} \chi_{it}, \nonumber \\
			&&\nonumber \\
			&\textrm{Model 3.4:}&\log r_{itI} = \alpha_I + \delta \kappa_{i} + v_i \gamma_{tI} + \varrho_t \chi_{it}, \nonumber \\
			&&\log r_{itM} =\alpha_M + \frac{1}{\delta} \kappa_{i} + u_i + \gamma_{tM} + \frac{1}{\varrho_t} \chi_{it}. \nonumber
		\end{eqnarray}
		
		\begin{table}[b!]
			\caption{\label{tabA2} Fitted models and their performance in terms of the selection criteria for each cancer site.}
				\begin{tabular}{cllrrrccrrr}
					\toprule
					&&&\multicolumn{3}{c}{Pancreatic cancer}&&&\multicolumn{3}{c}{Leukaemia cancer}\\
					\midrule
					&&& DIC & WAIC & LS &&& DIC & WAIC & LS\\
					\midrule
					\multicolumn{11}{l}{\bf Shared spatio-temporal interactions}\\
					\multicolumn{11}{l}{\hspace*{0.2cm} $\boldsymbol{l=T}$}\\
					&Model 3.1c& Type I& 16791 & 16546 & 8299 && Type IV& 17050 & 17071 & 8577\\
					&Model 3.2c& Type I& 16725 & 16509 & 8286 && Type IV& 17018 & 17023 & 8555\\
					&Model 3.3c& Type I& 16712 & 16503 & 8282 && Type IV& 16992 & 16988 & 8539\\
					&Model 3.4c& Type I& 16742 & 16542 & 8302 && Type IV& 16996 & 16995 & 8540\\
					\bottomrule
				\end{tabular}
			\end{table}
			
			As mentioned in the main paper, we consider different numbers of scaling parameters constructing alternative versions of the same model. Models with subindex a indicate $l=1$,  b indicates $l=3$ or $l=7$ scaling parameters and, c represent the models where the number of scaling parameter is different for each time period, i.e., $l=T$. Models a and b, along with their respective results, are presented in the main paper. Here, we focus on models c and their results.  For models c, we choose the extreme case where the model has a distinct parameter for each time period, denoted as $l = T$ (representing the most flexible model). \autoref{tabA2} presents the model selection criteria for models c. Comparing the criteria values of models c with the results in Table~5 of the main paper, it is evident that models c outperform the multivariate models with specific interactions for each health outcome. For pancreatic cancer, models c show worse criteria values. However, for leukemia, Model~3.3c demonstrates better WAIC values than Model~3.2b, which was selected in the main paper. Despite this, we choose Model~3.2b for analyzing leukemia data due to its simplicity.

		\newpage
		\section{Main results}\label{AppendixD}
		
		This section shows additional figures that were discussed but not included in the main paper due to space limitations.
		
		\begin{figure}[h!]
			\begin{center}
				\scalebox{0.25}{\includegraphics[page=1]{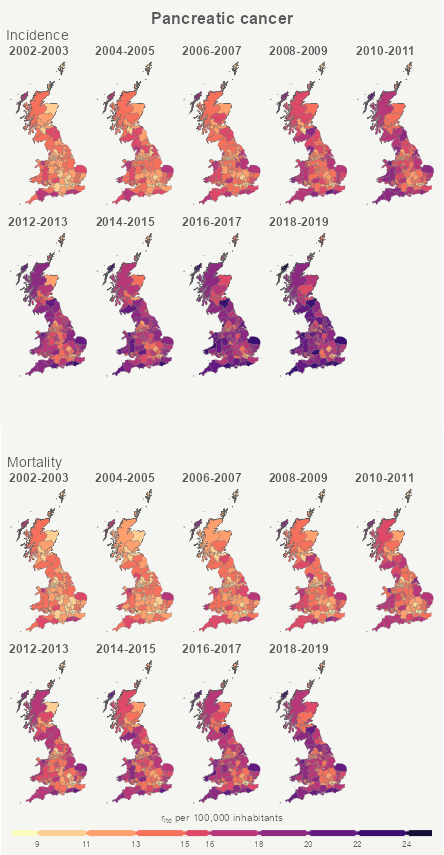}}
			\end{center}
			\caption{\label{figA2} Posterior medians of the evolution of the geographical distribution of rates per 100,000 inhabitants for pancreatic cancer.}
		\end{figure}
		
		\begin{figure}[h!]
			\begin{center}
			\scalebox{0.34}{\includegraphics[page=1]{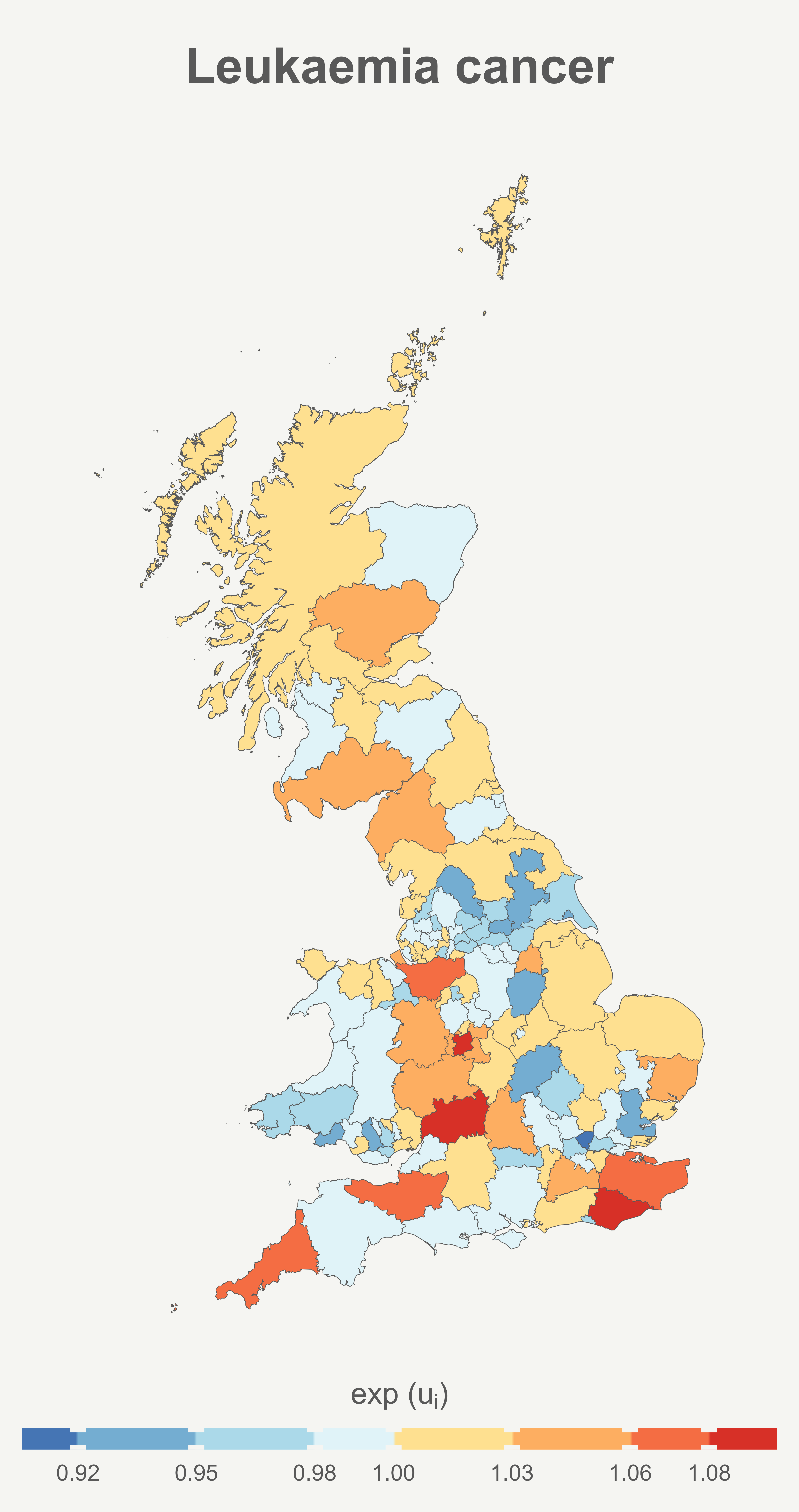}}
			\end{center}
			\caption{\label{figA1} Posterior median of the spatially unstructured random effect ($exp (u_{i})$) for leukaemia mortality.}
		\end{figure}
		
		\begin{figure}[h!]
			\begin{center}
				\scalebox{0.25}{\includegraphics[page=1]{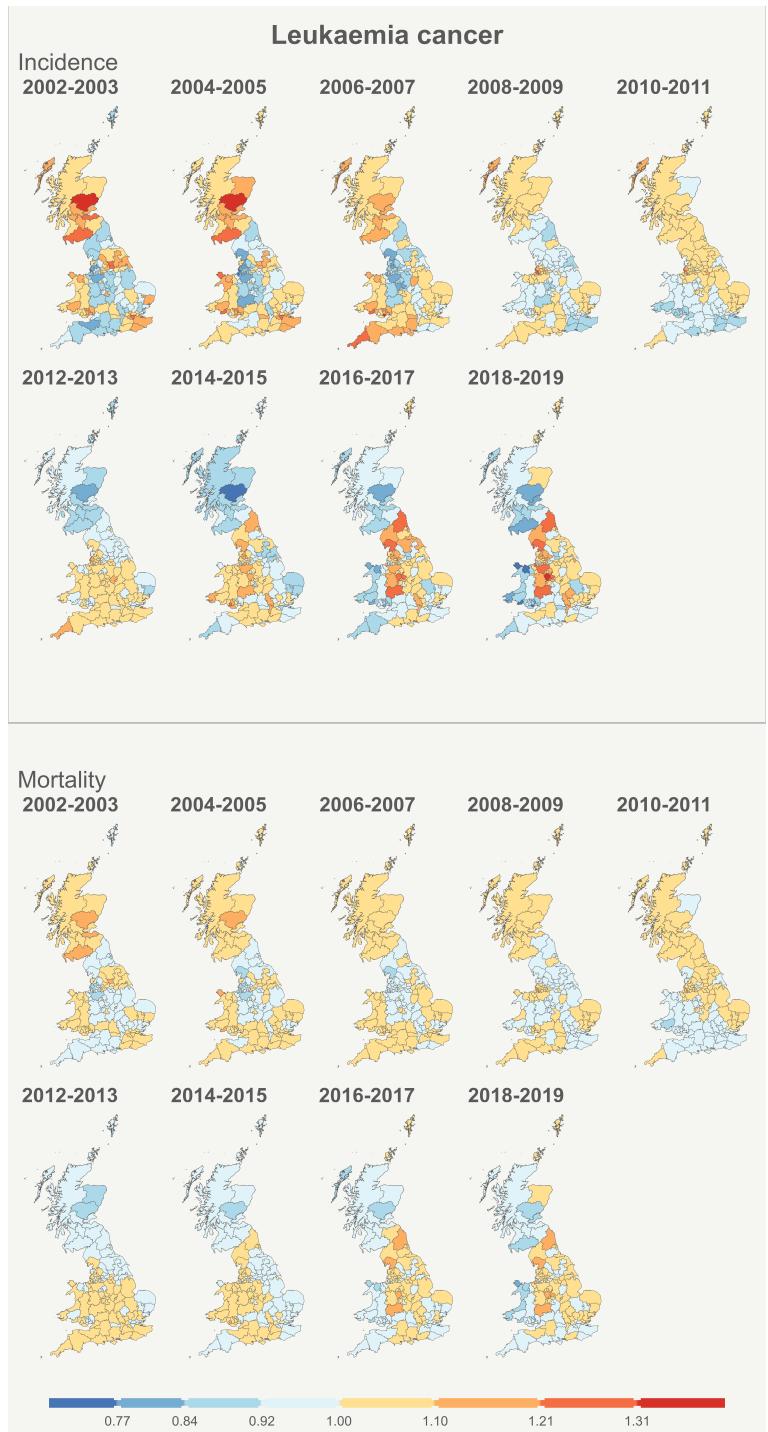}}
			\end{center}
			\caption{\label{figA1*} Posterior median of the spatio-temporal effect $(\exp (\varrho_{t}\chi_{it}))$ and $(\exp\frac{1}{\varrho_{t}}\chi_{it}))$ for leukaemia incidence and mortality respectively.}
		\end{figure}

		\begin{figure}[h!]
			\begin{center}
				\scalebox{0.25}{\includegraphics[page=1]{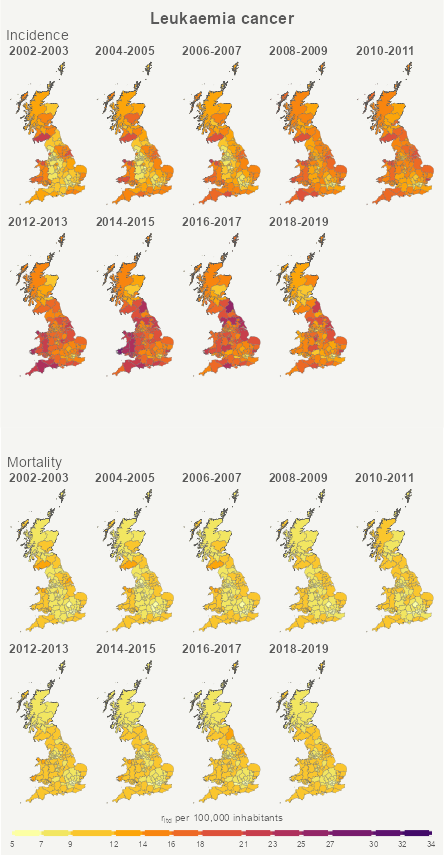}}
			\end{center}
			\caption{\label{figA3} Posterior median of the evolution of the geographical distribution of rates per 100,000 inhabitants for leukaemia.}
		\end{figure}
		
	\end{appendices}
	
\end{document}